\documentclass[preprint]{imsart}
\startlocaldefs

\usepackage[]{graphicx}
\usepackage{subfig}
\usepackage{epstopdf}
\usepackage{epsfig}
\usepackage{tabularx}
\usepackage{caption}
\usepackage[]{color}
\usepackage{xcolor}
\makeatletter
\def\maxwidth{ %
  \ifdim\Gin@nat@width>\linewidth
    \linewidth
  \else
    \Gin@nat@width
  \fi
}
\makeatother

\definecolor{fgcolor}{rgb}{0.345, 0.345, 0.345}
\newcommand{\IP}[1]{\left<#1\right>}

\usepackage{framed}
\makeatletter
 {\par\unskip\endMakeFramed%
 \at@end@of@kframe}
\makeatother

\definecolor{shadecolor}{rgb}{.97, .97, .97}
\definecolor{messagecolor}{rgb}{0, 0, 0}
\definecolor{warningcolor}{rgb}{1, 0, 1}
\definecolor{errorcolor}{rgb}{1, 0, 0}
\usepackage{alltt}

\usepackage[utf8]{inputenc}
\usepackage{amsmath}
\usepackage{amsfonts}
\usepackage{amssymb}
\usepackage[left=2cm,right=2cm,top=2cm,bottom=2cm]{geometry}
\usepackage{algorithm,algorithmicx,algpseudocode}
\usepackage{hyperref}
\usepackage{tikz}
\usepackage{graphicx}
\usepackage{bm}

\endlocaldefs

\begin{document}

\begin{frontmatter}

% "Title of the paper"
\title{A Geometric View of Posterior Approximation}
\runtitle{A Geometric View of Posterior Approximation}

\begin{aug}
 \author{\fnms{Tian} \snm{Chen,}\thanksref{t1}}
%\thankstext{t1}{Thanks to somebody} 
\author{\fnms{Jeffrey} \snm{Streets,}\thanksref{t2}}
\and
\author{\fnms{Babak} \snm{Shahbaba}\thanksref{t3, m1}\corref{Babak Shahbaba}\ead[label=e3]{babaks@uci.edu}}

\thankstext{t1}{Department of Statistics, UCI, tianc2@uci.edu}
\thankstext{t2}{Department of Mathematics, UCI, jstreets@math.uci.edu}
\thankstext{t3}{Department of Statistics, UCI, babaks@uci.edu}
\thankstext{m1}{Corresponding author}

\runauthor{Chen et~al.}

%\affiliation{Department of Statistics\thanksmark{m1} and Department of Mathematics\thanksmark{m2} \\ UC Irvine}
%\address{Department of Statistics\printead{tianc2@uci.edu}\\
%\phantom{E-mail:\ }\printead*{jstreets@math.uci.edu}}
\end{aug}

\begin{abstract}
Although Bayesian methods are robust and principled, their application in practice could be limited since they typically rely on computationally intensive Markov Chain Monte Carlo algorithms for their implementation. One possible solution is to find a fast approximation of posterior distribution and use it for statistical inference. For commonly used approximation methods, such as Laplace and variational free energy, the objective is mainly defined in terms of computational convenience as opposed to a true distance measure between the target and approximating distributions. In this paper, we provide a geometric view of posterior approximation based on a valid distance measure derived from ambient Fisher geometry. Our proposed framework is easily generalizable and can inspire a new class of methods for approximate Bayesian inference. 
\end{abstract}

%\begin{keyword}[class=MSC]
%\kwd[Primary ]{}
%\kwd{}
%\kwd[; secondary ]{}
%\end{keyword}

%\begin{keyword}
%\kwd{Fisher Geometry}
%\kwd{Variational Bayes}
%\end{keyword}

\end{frontmatter}

\section{Introduction}
In this paper, we are interested in approximating $p_{Z \vert X}(z\vert x)$, where $Z$ denotes model parameters or latent variables with prior distribution $p_Z(z)$ and $X$ denotes the observed data \cite{Bishop2007}. Inference regarding $Z$ typically involves integrating functions over the posterior, 
\begin{eqnarray}
E_{Z \vert X} (g(z))& = & \int g(z)p_{Z \vert X}(z\vert x) dz
\end{eqnarray}
For instance, $g(z) = z$ if we are interested in estimating the posterior mean. Unfortunately, the integration problem in Bayesian inference is not analytically tractable in most cases. To address this issue, we could use Markov Chain Monte Carlo (MCMC) algorithms by simulating large samples from intractable posterior distributions and using these samples to approximate the above integral. However, MCMC algorithms tend to be computationally intensive, especially for large scale problems. Although many methods have been proposed in recent years to improve computational efficiency of MCMC algorithms (see for example, \cite{neal96a, neal93, PMCMC, mykland95, ProppWilson96, roberts97, gilks98, warnes01, freitas01, brockwell06, neal11, neal05, neal03, Beal03, MollerPettittBerthelsenReeves06, AndrieuMoulines06, KuriharaWellingVlassis06, cappe08, craiu09, wellingUAI09, GelfandMaatenChenWelling10, douc07, douc11, wellingTeh11, ZhaSut2011a, ahmadian11, girolami11, hoffman11, shahbabaSplitHMC, beskos11, calderhead12, lan14}), extending these methods to high dimensional and complex distributions remains a challenge. 

Here, we focus on an alternative family of methods based on deterministic approximation of posterior distribution to cope with intractable problems in Bayesian inference. These methods aim at finding an approximate, but tractable, distribution to replace the exact posterior distribution in order to make statistical inference easier. For example, Laplace's method approximates posterior distribution by a Gaussian distribution with the mean set at the mode of the posterior distribution and the covariance set to the second derivative of the log posterior density evaluated at the mode. While this approach is quite easy to implement, in most cases it could only provide good local approximation around the mode; that is, it could fail to capture global features of the posterior distribution \cite{Bishop2007}. 

An alternative approach is to use variational Bayes methods that approximate the posterior distribution by a much simpler distribution, $p'_Z(z)$, which is assumed to belong to a specific family of models. A divergence is then specified to quantify the dissimilarity between $p_{Z \vert X}(z \vert x)$ and $p'_Z(z)$. An optimal $p'_Z(z)$ is chosen from the family of valid distributions by minimizing the divergence. 

The variational free energy method (VFE) developed by Feynman and Bogoliubov \cite{mackay03} uses the relative entropy, usually referred to as Kullback-Leibler divergence (KL-divergence), as the measure of dissimilarity between the target and approximating distribution. Consider the following decomposition of the log marginal likelihood:
\begin{eqnarray}
\log p_X(x) & = & \int p'_Z(z) \log \dfrac{p_{X,Z}(x,z)}{p'_Z(z)}dz - \int p'_Z(z) \log \dfrac{p_{Z \vert X}(z \vert x)}{p'_Z(z)}dz \\
& = & \mathcal{L}(p'_Z) + KL(p'_Z \Vert p_{Z \vert X})
\end{eqnarray}
Because KL divergence is non-negative, $\mathcal{L}(p'_Z)$ serves as a lower bound for $\log p_X$.  $\mathcal{L}(p'_Z)$ is often referred to as the (negative) variational free energy.  Because $\log p_X$ is fixed with respect to $p'_Z$, minimizing KL-divergence is equivalent to maximizing the lower bound $\mathcal{L}(p'_Z)$; that is, the optimal approximation distribution can be obtained by  
\begin{eqnarray}
{p'}_Z^* = \arg \min \limits_{p'_Z \in P'} KL(p'_Z \Vert p_{Z \vert X}) = \arg \max \limits_{p'_Z \in P'} \mathcal{L}(p'_Z) 
\end{eqnarray}
where $P'$ is the set of all valid densities. %Here, ${p'}_Z^*$ is referred to as the Q-VB approximation.   
The purpose of restricting $p'_Z$ to $P'$ is to make the integration over $p'_Z$ tractable and to simplify the optimization problem. In practice, it is common to assume that $p'_Z(z)$ is factorizable.
 
Note that KL-divergence is not symmetric in general: $KL(p' \Vert p)$ is not the same as its reverse $KL(p \Vert p')$. While methods based on variational free energy typically use $KL(p' \Vert p)$, an alternative method, known as ``expectation propagation'' \cite{Minka2001}, uses the reverse KL: $KL(p \Vert p')$. In this case, by restricting $p'_Z(z)$ to the exponential family, minimization of the reverse KL is  simply a moment matching algorithm; that is, by setting the expectation of sufficient statistics of $p'_Z$ equal to that of $p_{Z \vert X}$, we minimize the reverse KL-divergence. However, the results from direct optimization can be highly inaccurate. Expectation propagation views the joint distribution as a product of factors: $p_{X,Z}(x,z) = \prod_{i=0}^n t_i(z)$ where $t_0(z)$ represents the prior and $t_i(z)$ corresponds to the likelihood of data point $x_i$. The approximating distribution $p'_Z(z)$ is then also assumed to be factorizable: $p'_Z(z)) \propto \prod_{i=0}^n \tilde{t}_i(z)$. Each $\tilde{t}_i(z)$ represents an approximating function of $t_i(z)$. The algorithm starts by initializing $\tilde{t}_i(z)$ for $i = 1, \ldots, n$. Then, given $\tilde{t}_{i \neq j}(z)$, each $\tilde{t}_j(z)$ is updated iteratively by moment matching between $p'_Z(z)$ and $t_j(z) \prod_{i\neq j} \tilde{t}_i(z) $.
 
It is worth noting that above divergence measures are special cases of a family of divergence measure known as $\alpha$-divergence (\cite{Zhu1995} \cite{Minka2005}),
\begin{eqnarray}
D_{\alpha} (p \Vert p') = \dfrac{\int \alpha p(x) +(1-\alpha) p'(x) - p(x)^{\alpha}p'(x)^{1-\alpha} dx}{\alpha(1-\alpha)}, \quad \alpha \in (-\infty, \infty)
\end{eqnarray}
The variational free energy method is a special case when $\alpha \rightarrow 0$ so we have $\lim_{\alpha \rightarrow 0} D_{\alpha} (p \Vert p') = KL(p' \Vert p)$, while the reverse KL divergence corresponds to $\lim_{\alpha \rightarrow 1} D_{\alpha} (p \Vert p') = KL(p \Vert p')$. Setting $\alpha = 0.5$, we obtain a symmetric measure $D_{0.5}(p \Vert p') = 2 \int (\sqrt{p(x)} - \sqrt{p'(x)})^2 dx = 4H^2(p,p')$ where $H(p,p')$ represents the Hellinger distance defined as
\begin{eqnarray}
H(p,p') = \left( \frac{1}{2} \int (\sqrt{p(x)}-\sqrt{p'(x)})^2 dx  \right)^{\frac{1}{2}} = \dfrac{1}{\sqrt{2}} \Vert \sqrt{p(x)}-\sqrt{p'(x)} \Vert_2 
\end{eqnarray}

Note that $\alpha$-divergence is not symmetric except for the case of the Hellinger distance ($\alpha = 0.5$), which is rarely used in variational Bayes methods. Most existing variational methods, such as variational free energy and expectation propagation, do not use a real metric for quantifying the approximation error. In contrast to these existing methods, our proposed method, called Geometric Approximation of Posterior (GAP), is based on the \textit{ambient Fisher information metric} that uses a true distance measure, which we call \textit{spherical Fisher distance}. Theoretically, this method provides a novel view of approximate Bayesian inference from the perspective of statistical geometry. Practically, it is a promising method that has the potential to overcome the shortcomings of existing methods. More specifically, unlike MCMC methods, our method does not require computationally intensive simulations. Compared to existing approximation methods, it relies on a true metric and is more flexible in terms of defining the approximating family of distributions. 

This paper is organized as follows. In the following section, we present our method based on the spherical Fisher distance. In Section \ref{sec:illustrations}, we illustrate this approach using simple examples. Finally, we discuss several future directions in Section \ref{sec:discussion}. 
 
\section{Methods}\label{sec:methods}
In this section, we present our geometry-based method for approximating posterior distributions. First, we provide a brief overview on geometry of statistical models in general. Next, we discuss ``Ambient'' Fisher geometry (AFG), which is a particular view of statistical models first observed by \cite{A.P1977} (cf. \cite{amari00,Lebanon2005}), but has remained relatively unknown in the statistics community. Finally, we show how this geometric view of statistical models can be used to approximate posterior distributions.  

%KL-divergence is not a real distance. However, we develop a distance (referred to as the spherical Fisher distance) to measure the dissimilarity between $P(Z \vert X)$ and $Q(Z)$. We first introduce and justify for this distance. Then we will introduce an algorithm using projection on hemisphere plane to find the specific $\theta$ which minimizes the spherical Fisher distance.

\subsection{The Fisher Metric}
Let $(M,g)$ be a smooth Riemannian manifold and let ${\mathcal P}$ denote the space of probability
distributions on $M$.  We will use the volume form $dV_g$ to identify distributions with smooth functions which integrate to $1$ against this volume form.  We can interpret a model $\Theta$ as a map from an open
set in some parameter space $U \subset \mathbb R^D$%\footnote{Do you really want $\mathbb R^p$ here?  It's too many $p$'s.  Can we use $n$?  Does this $\mathbb R^p$ appear elsewhere?} 
to ${\mathcal P}$, i.e.
\begin{align*}
 \Theta : U \to {\mathcal P}, \qquad (\theta_1,\dots,\theta_D) \to p_{\theta}.
\end{align*}
 Denote the associated set of distributions by $S = \{p_{\theta}   \vert \theta = [\theta_1,\cdots, \theta_D] \}$, which lies in $L^1$ space and is a subset of $\mathcal P$. $S$ is often regarded as an D-dimensional manifold endowed
with a Riemannian metric using  Fisher information matrix. By introducing a Riemannian metric (i.e. a local inner product on the tangent space at each point) on the manifold, we can derive many geometric notions such as length of curves, geodesic and distance. The Fisher metric is a Riemannian metric defined by Fisher information matrix:
\begin{align*}
g^{\Theta}_{F} \left(\frac{\partial}{\partial \theta_i}, \frac{\partial}{\partial
\theta_j} \right)_{p} = \int_M \left( \partial_{\theta_i} \log p_{\theta} \right) \left(
\partial_{\theta_j} \log p_{\theta} \right) p_{\theta} dV_g.
\end{align*}
where $\frac{\partial}{\partial \theta_i}, \frac{\partial}{\partial
\theta_j}$ are the $i^{th}$ and $j^{th}$ basis vectors of the tangent space at point $p_{\theta}$.

\subsection{Ambient Fisher geometry}
 
In what follows we give a brief summary of ``Ambient'' Fisher geometry (AFG). This
point of view has appeared in the literature, although is not very
well-known. Our particular viewpoint was first observed in \cite{A.P1977} (cf.
\cite{amari00,Lebanon2005}). 

The Fisher information metric can be interpreted as the Riemannian metric induced by an ambient metric
on the infinite dimensional manifold ${\mathcal P}$.
To do this we observe that for a given $p \in {\mathcal P}$, the tangent space
can be identified with
\begin{align*}
 T_p {\mathcal P} := \left\{ \phi \in C^{\infty}(M)\ |\ \int_M \phi dV_g = 0 \right\},
\end{align*}
which arises by differentiating the unit mass condition on probability
distributions.  We can then define the ambient Fisher metric on ${\mathcal P}(X)$ by
\begin{align} \label{fisher}
 g_F^{{\mathcal P}}(\phi,\psi)_{p} := \int_M \frac{\phi \psi}{p} dV_g.
\end{align} 
A direct calculation shows that for a model $\Theta : U \to {\mathcal P}$, we have
$\Theta^* g_F^{\mathcal P} = g_F^{\Theta}$.  In other words, the Riemannian geometry induced
by the ambient metric on the image of the embedding of the model into ${\mathcal P}$ is
the usual Fisher metric.

Our goal is to use the ambient geometric  structure of ${\mathcal P}$ to better understand
properties of specific models $\Theta$.  As it turns out, many geometric
properties become clearer when one changes point of view and interprets
probability distributions as the unit sphere in the $L^2$ metric as opposed to
the $L^1$ metric.  Specifically, let $\mathcal Q = \{ q : M \to \mathbb R | \int_M q^2
dV_g = 1 \}$.  We can endow the space of $L^2$ functions on $M$ with the usual
flat inner product, although now interpreted as a Riemannian metric. This induces an
inner product on $\mathcal Q$, called $g_F^{\mathcal Q}$, which is in direct analogy with the
geometry inherited by the unit sphere in an ambient Euclidean space.  Moreover,
direct calculations show that the map $\mathcal S : \mathcal Q \to \mathcal P$ defined by $\mathcal S(q) = q^2$
is a Riemannian isometry, i.e. $\mathcal S^* g_F^{\mathcal P} = g_F^{\mathcal Q}$.  Thus it is
equivalent to work in the space $\mathcal Q$ instead of $\mathcal P$, which we will now do
exclusively.

Using the picture of $\mathcal Q$ as the unit sphere of the space of $L^2$ functions,
we can formally derive many basic equations which are fundamental in
understanding the ambient Fisher geometry.  For instance, we can explicitly
solve for geodesics in $\mathcal Q$.  First, given $q_0 \in \mathcal Q$ and $f \in T_q \mathcal Q$ a
unit tangent vector, the geodesic with initial value $q_0$ and initial unit norm velocity
$f$ exists on $(-\infty,\infty)$ and takes the form
\begin{align} \label{velocitygeod}
 q_t = q_0 \cos t + f \sin t.
\end{align}
The obvious $2\pi$-periodicity is no surprise, as this curve corresponds to a
great circle in the infinite dimensional sphere $\mathcal Q$.  Also, given $q, q' \in
\mathcal Q$, the geodesic connecting them takes the form
\begin{align} \label{twopntgeod}
 q_t = q \cos t + \frac{q' - q \left< q',q \right>}{\left| q' - q \left< q',q \right> \right|} \sin t
\end{align}
Observe that this is well-defined if and only if $q' \neq \pm q$.  This
makes sense as there is no canonical direction to point in to head from the
north pole to the south pole.  In this exceptional case one can obtain a
geodesic connecting $q$ and $q'$ by choosing an arbitrary initial velocity $f$
and using (\ref{velocitygeod}).  Moreover, a direct integration using
(\ref{twopntgeod}) shows that the distance between two point $q,q'$ is the
``arccosine'' distance, i.e.
\begin{align} \label{Qdistance}
 d^{\mathcal Q}_F(q,q') = \arccos \int_M q q' dV_g,
\end{align}
which we refer to as spherical Fisher distance in this paper. But since the map $\mathcal S : \mathcal Q \to \mathcal P$ is an isometry, we have the distance between two distributions $p = q^2$, $p' = q'^2$ $\in \mathcal{P}$ defined as:
\begin{eqnarray}
d_{SF}(p,p') \equiv d_{F}^{\mathcal{P}} (p,p') = d_F^{\mathcal{Q}}(q, q') = \arccos \int_M \sqrt{pp'} dV_g
\end{eqnarray}

Notice that the distance associated with the usual flat inner product in the ambient ``Euclidean space'' (i.e. the space of $L^2$ functions) is: 
\begin{eqnarray}
d_H(p,p') = \left(  \int_M (\sqrt{p}-\sqrt{p'})^2 dV_g  \right)^{\frac{1}{2}},
\end{eqnarray} 
which is directly related to the Hellinger distance. In contrast, spherical Fisher distance is the distance associated with the inner product on the ``unit sphere'' manifold $\mathcal{Q}$ (the space of square roots of probability distributions) induced by the usual flat inner product. Although the metric used in our method is different from the Hellinger distance, the two metrics are related in that minimizing spherical Fisher distance is equivalent to minimizing the Hellinger distance between the target and approximating distributions. Geometrically, however, using the spherical Fisher distance is more justifiable and can be optimized more smoothly.

\subsection{Variational Bayes using AFG}\label{VB-AFG}
In spite of the difficulty in visualizing a class of distributions (e.g., normal distribution) on the ``unit sphere" $\mathcal{Q}$, we could still make use of this idea to approximate complicated distributions through variational methods: after we specify a class of distributions, our task is to find a member of this family with the shortest distance to the target distribution (e.g., posterior distribution). That is, we approximate the target distribution by $p'_Z(z)$, i.e., a member of the assumed family of distributions, by minimizing the spherical Fisher distance to $p_Z(z)$. Notice that unlike KL, the spherical Fisher distance used in our method is based on a true metric. In what follows, we illustrate this idea using a simple problem with analytical solution.

Consider a Gaussian model, $x\sim N(\mu, \tau^{-1})$, with unknown mean, $\mu$, and variance, $\tau^{-1}$ (here, $\tau$ is the precision parameter). Although the posterior distribution is not tractable in general, it is possible to simplify the problem and find an analytical form for the posterior distribution by connecting the prior variance of $\mu$ to the variance of data as follows:
\begin{eqnarray*}
\mbox{Prior:} & \mu \vert \tau \sim N(\mu_0, (\lambda_0\tau)^{-1}) \\
 & \tau \sim Gamma(\alpha_0, \beta_0)  
\end{eqnarray*}
This prior is known as the Normal-Gamma distribution. In this case, given $n$ observed values for $x$, the posterior distribution has a closed form: $$(\mu, \tau |x) \sim N(\mu_N^*, (\lambda_N^* \tau)^{-1})Gamma(\alpha_N^*, \beta_N^*),$$ 
where
\begin{eqnarray*}
\mu_N^* & = & \dfrac{n\bar{x}+\lambda_0\mu_0}{n+\lambda_0}\\
\lambda_N^* & = & \lambda_0+ n \\
\alpha_N^* & = & \alpha_0+\dfrac{n}{2}\\
\beta_N^* & = & \beta_0 + \dfrac{S}{2} + \dfrac{n\lambda_0(\bar{x}-\mu_0)^2}{2(n+\lambda_0)}
\end{eqnarray*}
In Appendix \ref{illustExample1}, we show that if we limit our approximating distributions also to the Normal-Gamma family:
\begin{eqnarray*}
 \mu \vert \tau & \sim & N(\mu_N, (\lambda_N\tau)^{-1}) \\
 \tau & \sim & Gamma(\alpha_N, \beta_N)
\end{eqnarray*}
then minimizing spherical Fisher distance with respect to $(\mu_N, \lambda_N, \alpha_N, \beta_N)$ leads to the exact same posterior distribution shown above. That is, by minimizing the spherical Fisher distance between the true posterior $p$ and the approximating distribution $p'$, the optimal $p'$ is exactly $p$.

\subsection{Gradient Descent Algorithm}
\begin{figure}[t]
\centering
\includegraphics[width=0.48\textwidth]{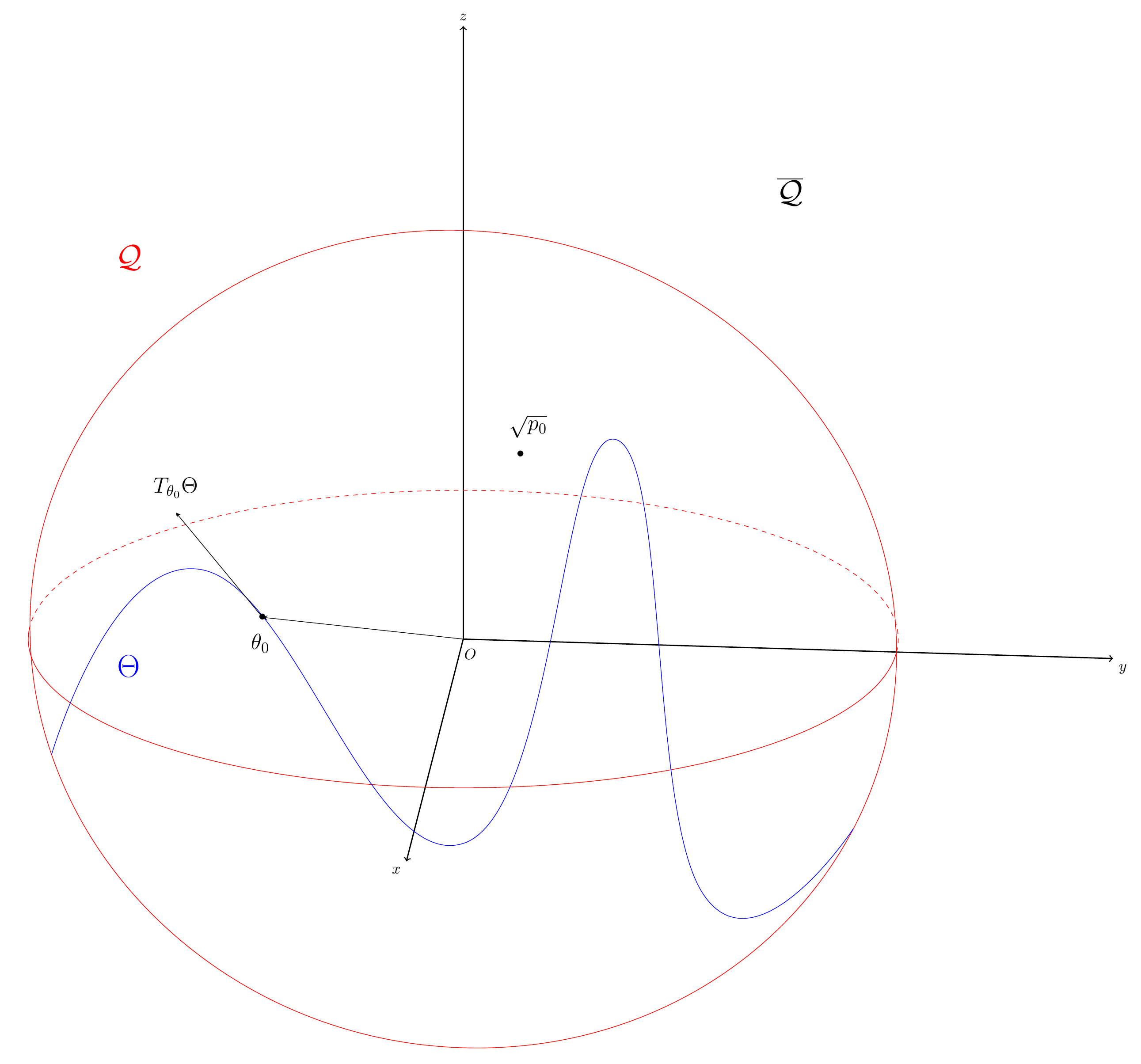}
\caption{A schematic representation of our method. $\Theta$ represents the class of approximation distributions. We start from an arbitrary point $\theta_0$ on $\Theta$. Its tangent space with respect to the manifold $\Theta$ is denoted as $T_{\theta_0}\Theta$. We move from $\theta_0$ to a new point on $\Theta$ directed by the negative gradient vector of the distance function. This is the same direction as the projection of $\sqrt{p_0}$ onto $T_{\theta_0}\Theta$.}\label{fig:algm}
\end{figure}

In general, there is no analytical solution for the optimization problem in our method. To address this issue, we develop a gradient-descent optimization algorithm to minimize the distance function. Suppose $p_0$ is an intractable target distribution (here, posterior distribution) that we want to approximate using a parametric model from $\Theta$. We start from an arbitrary point $\theta_0 \in \Theta$ and improve the approximation via a modified gradient descent in $\mathcal Q$. Note that $\sqrt{p_0} \in \mathcal Q$ and the model $\Theta$ is naturally embedded in $\mathcal Q$. Using (\ref{Qdistance}), we can calculate the gradient of the distance function.  In particular, given a single parameter family $\theta_t \in \Theta$ with derivative $\dot{\theta}$, a direct calculation shows that the directional derivative takes the following form:
\begin{align} \label{gradd}
\nabla_{\dot{\theta}} d(\theta, \sqrt{p_0}) = - \frac{\IP{\dot{\theta},\sqrt{p_0}}}{\sqrt{ 1 - \IP{\theta,\sqrt{p_0}}^2}}.
\end{align}
Because our possible choices of $\dot{\theta}$ are restricted to $T_{\theta} \Theta$, it is clear that this directional derivative will be minimized by projecting the vector $\sqrt{p_0}$ onto $T_{\theta} \Theta$,
\begin{eqnarray*}
\mbox{proj}_{T_{\theta} \Theta}{\sqrt{p_0}} & = & \sum w_i \langle w_i, \sqrt{p_0} \rangle,
\end{eqnarray*}
where $\{ w_i \}$ are orthonormal basis for $T_{\theta} \Theta$.  Ultimately combining this with (\ref{gradd}) yields the negative gradient vector
\begin{align} \label{vdef}
\overrightarrow{v_0} = \frac{\mbox{proj}_{T_{\theta} \Theta}{\sqrt{p_0}}}{\sqrt{1 - \IP{\theta, \sqrt{p_0}}^2}} = \frac{\sum w_i \IP{w_i,\sqrt{p_0}}}{\sqrt{1 - \IP{\theta, \sqrt{p_0}}^2}}.
\end{align}
Therefore, to find an optimal solution for an arbitrary class of models, $\Theta$, we start from an initial point $\theta_0$ on $\Theta$ and follow these steps (Figure \ref{fig:algm}):

\begin{minipage}[t]{0.9\textwidth}
\begin{description}
\vspace{12pt}
\item[Step 1] Given $\theta_0$, compute $v_0$ as in (\ref{vdef}).
\vspace{16pt}
\item[Step 2] Move from $\theta_0$ to $\theta_1$ guided by $\overrightarrow{v_0}$ while confined to $\Theta$. For this, ideally we could follow the geodesic of $\Theta$ with $\theta_0$ as the initial position and $\overrightarrow{v_0}$ as the initial velocity to update the parameters. However, because of the difficulty in obtaining such geodesics in general cases, we can instead follow an approximate path. To this end, we set $\overrightarrow{v_0} = \sum \limits_{i=1}^D \alpha_i w_i$ and update the parameters separately in each direction: $\theta_1^i = \theta_0^i+\epsilon \alpha_i$ for $i = 1, \ldots, D$, where $\epsilon$ is the step size.
 Iterate the above steps until the updated values of parameters remain close to the current values (i.e., current negative gradient vector $\overrightarrow{v} \approx 0$) or the distance between the target and approximating distributions falls below a predefined threshold.
 \vspace{12pt}
\end{description}
\end{minipage}

\noindent
We iterate through the above steps to obtain the closest point on $\Theta$ to $\sqrt{p_0}$.  
  
%\begin{figure}[htbp]
%\centering
%\subfloat[]{\label{fig:subfig_a}
%  \includegraphics[width=0.48\textwidth]{Algorithm1.pdf}
%  }
%\subfloat[]{ \label{fig:subfig_b}
%  \includegraphics[width=0.48\textwidth]{Algorithm4.pdf}
%  }
%  \caption{One iteration of gradient descent \textcolor{red}{need new picture}}
%
%\end{figure}

\subsection{Gaussian approximation}
\label{sec:algm1}
As discussed above, in practice we usually define a simple class of models to approximate target distributions. Here, we discuss how any arbitrary target distribution can be approximated by a multivariate Gaussian distribution using our method. The resulting algorithm is based on the matrix representation of Gram-Schmidt process. The full details of the procedure can be found in Appendix \ref{appd: algm1}.

Suppose $\Theta$ represents the family of Gaussian models $N(z \vert \mu, \Sigma)$. Then any point on $\Theta$ can be expressed as: $$p(z \vert \mu, \Sigma) =  (2\pi)^{-\frac{D}{2}} \vert \Sigma \vert^{-\frac{1}{2}}  \exp\left(-\frac{1}{2}( z - \mu)^T{\Sigma}^{-1}(z - \mu) \right),$$ with the corresponding square root of the density, $$q(z \vert \mu, \Sigma) = (2\pi)^{-\frac{D}{4}} \vert \Sigma \vert^{-\frac{1}{4}} \exp\left(-\frac{1}{4}( z - \mu)^T{\Sigma}^{-1}(z - \mu) \right)$$. 
Since $\Sigma$ is constrained to be positive definite, we use its Cholesky decomposition, $\Sigma = LL^T$, and minimize the distance with respect to the lower triangular matrix, $L$, with unconstrained parameterization. Also, we sometimes express the covariance parameters in a vector form for simplicity. This is achieved by using the $vech$ operator which vectorizes a matrix column-wise, while excluding the upper part of the matrix \cite{fackler2005notes}:$vech(L) = l$. 

In order to obtain an orthonormal basis of the tangent space at any point on $\Theta$, consider the push-forwards of basis vectors $\{ \frac{\partial}{\partial \theta_i} \}$ with respect to the map from parameter space to root distribution space: $\{ \frac{\partial q}{\partial \theta_i} \}$. Thus, starting from an initial point on $\Theta$, $\theta_0 =  (\mu_0, L_0)  \equiv q(z \vert \mu_0, \Sigma_0) \equiv q_0$, the orthonormal basis $\{ w_i \} $ of the tangent space $T_{\theta_0}\Theta$ can be obtained by orthonormalizing the following basis:
\begin{eqnarray*}
 v_{\mu}  =   \dfrac{\partial q}{\partial \mu} \Big \vert_{\mu = \mu_0, L = L_0} & = & q(z \vert \mu_0, \Sigma_0)\dfrac{1}{2} (z-\mu_0)^T \Sigma_0^{-1}, \quad ({1 \times D} \mbox{ vector}) \\
v_l =   \dfrac{\partial q}{\partial vech(L)} \Big \vert_{\mu = \mu_0, L = L_0} & = &   q(z \vert \mu_0, \Sigma_0)  \left[ -\dfrac{1}{4} vec(\Sigma_0^{-T})^{T} +\dfrac{1}{4}((z-\mu_0)^T \otimes (z-\mu_0)^T)(\Sigma_0^{-T} \otimes \Sigma_0^{-1})\right] \\
& & \left[I+T_{D,D}^T-R_D^T \right] \left[ (I_D \otimes L_0) T_{D,D} + (L_0 \otimes I_D) \right] S_D^T , \quad  (1 \times {\dfrac{D(D+1)}{2}} \mbox{ vector}) 
\end{eqnarray*}
Given $\{ w_i \} \equiv \left( \{w_{\mu_i} \}_{i=1}^D,\{{w_l}_i \}_{i=1}^{\frac{D(D+1)}{2}} \right)$, we have $$\overrightarrow{v_0} = \dfrac{\sum \limits_{i=1}^D \langle {w_\mu}_i, \sqrt{p_0}\rangle {w_\mu}_i + \sum \limits_{i=1}^{D(D+1)/2} \langle {w_l}_i, \sqrt{p_0}\rangle {w_l}_i}{\sqrt{1-\langle \theta_0, \sqrt{p_0}\rangle^2}}$$ Finally, we update the parameters as follows:
\begin{eqnarray*}
 \mu_i^{(t+1)} & = & \mu_i^{(t)}+\epsilon_{\alpha_i} \langle w_{\mu_i}, \sqrt{p_0}\rangle  / \sqrt{1-\langle \theta_0, \sqrt{p_0}\rangle^2} \quad  i = 1, \cdots , D\\
 l_i^{(t+1)} & = &l_i^{(t)}+\epsilon_{\beta_i} \langle {w_l}_i, \sqrt{p_0}\rangle  / \sqrt{1-\langle \theta_0, \sqrt{p_0}\rangle^2} \quad  i = 1, \cdots , \frac{D(D+1)}{2}
\end{eqnarray*} where $\epsilon_{\alpha_i}, \epsilon_{\beta_i} $ are stepsizes. Algorithm \ref{alg:innerMu} and \ref{alg:innerSigma2} show the steps to obtain $\langle w_{\mu_i}, \sqrt{p_0}\rangle$ and $\langle {w_l}_i, \sqrt{p_0}\rangle$ respectively. 

%\alglanguage{pseudocode}
\begin{algorithm}[t]
\caption{Obtaining $\langle w_{\mu_j}, \sqrt{p_0}\rangle$}
\label{alg:innerMu}
\begin{algorithmic}
\State Generate iid samples $z^{(t)}, t = 1, \cdots, T$ from $q^2 (z \vert \mu_0, \Sigma_0)$ %\Comment {Get $\langle w_{\mu_j}, \sqrt{p_0}\rangle$, $j = 1, \cdots, D$}
\State Calculate $a_i$: the mean of $\frac{\sqrt{p_0(z^{t})}}{q(z^{(t)} \vert \mu_0, \Sigma_0)}\frac{1}{2} \left[  (z^{(t)}-\mu_0)^T \Sigma_0^{-1} \right]_i $, $ t = 1, \cdots, T$ for each $i = 1, \cdots, D$
\State Let $A = \frac{1}{4}\Sigma_0^{-1}$
\For{ $j = 1 $ to $D$}
\State Obtain $A_j$ as the $j$th order leading principal submatrix of $A$
\State Calculate $D_j $ as the determinant of $A_j$
\State Calculate $M_{j,i}$ for each $i = 1,\cdots,j$, where $M_{j,i}$ is a minor of $A_j$
\State Calculate $\langle w_{\mu_j}, \sqrt{p_0}\rangle  = \frac{1}{\sqrt{D_{j-1} D_j}} \sum \limits^j_{i=1} (-1)^{j+i}  M_{j,i} a_i $
\EndFor 
\end{algorithmic}
\end{algorithm}

\begin{algorithm}[t]
\caption{Obtaining $\langle {w_l}_j, \sqrt{p_0} \rangle$}
\label{alg:innerSigma2}
\begin{algorithmic}
\State Pre-calculate $T_{D,D}, R_D, S_D$ as defined in Appendix \ref{appd: algm1}. 
\State Pre-calculate  $U_D = I+T_{D,D}^T-R_D^T$.
\\
\State Calculate $V_D = \left[ (I_D \otimes L_0) T_{D,D} + (L_0 \otimes I_D) \right] S_D^T  $
\State Calculate $E(W_D^TW_D)$ (Appendix \ref{appd: algm1}. First obtain $vec(\Sigma_0)$, $\Sigma_0^{-1}$, $vec( \Sigma_0^{-1} )$, $\Sigma_0 \otimes \Sigma_0$, $\Sigma_0^{-1} \otimes \Sigma_0^{-1}$. Permute $\Sigma_0 \otimes \Sigma_0$ to obtain $\left[ (\Sigma_0 \otimes \Sigma_0)_{ikjl} \right], \left[ (\Sigma_0 \otimes \Sigma_0)_{iljk} \right]$)
\\
\State Calculate $B =  V_D^TU_D^TE(W_D^TW_D)U_DV_D $
\\
\State Generate iid samples $z^{(t)}, t = 1, \cdots, T$ from $q^2 (z \vert \mu_0, \Sigma_0)$ %\Comment {Get $\langle w_{\mu_j}, \sqrt{p_0}\rangle$, $j = 1, \cdots, D$}
\State Calculate $b_i$: the mean of $\frac{\sqrt{p_0(z^{t})}}{q(z^{(t)} \vert \mu_0, \Sigma_0)} \left[ \left( -\dfrac{1}{4} vec(\Sigma_0^{-T})^{T} +\dfrac{1}{4}((z^{(t)}-\mu_0)^T \otimes (z^{(t)}-\mu_0)^T)(\Sigma_0^{-T} \otimes \Sigma_0^{-1}) \right) U_DV_D\right]_i$, $ t = 1, \cdots, T$ for each $i = 1, \cdots, \frac{D(D+1)}{2}$
\For{ $j = 1 $ to $\frac{D(D+1)}{2}$}
\State Obtain $B_j$ as the $j$th order leading principal submatrix of $B$
\State Calculate $E_j $ as the determinant of $B_j$
\State Calculate $N_{j,i}$ for each $i = 1,\cdots,j$, where $N_{j,i}$ is a minor of $B_j$
\State Calculate $\langle w_{l_j}, \sqrt{p_0}\rangle  = \frac{1}{\sqrt{E_{j-1} E_j}} \sum \limits^j_{i=1} (-1)^{j+i}  N_{j,i} b_i $
\EndFor 
\end{algorithmic}
\end{algorithm}

\section{Illustrations}
In this section, we evaluate our approximation method using three illustrative examples. We first start with a toy example, where we approximate a $t$-distribution with a normal distribution. Next, we use our method to find a normal approximation to the posterior distribution of parameters in a Bayesian logistic regression model. Our final example involves approximating a bimodal distribution, which is a mixture of two normals. 
\label{sec:illustrations}
\subsection{A toy example: approximating the $t(1)$ distribution}
For our first example, we use our method to find a normal approximation, $N(\mu, \sigma^2)$, to the $t$-distribution with 1 degree of freedom, $t(1)$. Although this is just a one-dimensional case of the procedure discussed in section \ref{sec:algm1}, we would like to elaborate it in more details here. For this problem, we have
\begin{alignat*}{3}
& \mbox{the square root density of $t$(1): } & \sqrt{p_0(x)} &  = \pi^{-\frac{1}{2}}(1+x^2)^{-\frac{1}{2}} \\
& \mbox{the square root density of } N(\mu, \sigma^2): \quad   & \sqrt{p(x \vert \mu, \sigma^2)} & =  (2\pi)^{-\frac{1}{4}}(\sigma^2)^{-\frac{1}{4}}\exp (-\dfrac{1}{4\sigma^2}(x-\mu)^2) \equiv  q(x \vert \mu, \sigma^2) 
\end{alignat*}
To obtain unconstrained parameterization, we update $\sigma$ ($ - \infty < \sigma < \infty$) instead of $\sigma^2$.  The basis for the tangent space $T_{\theta_0}\Theta$ are as follows:
\begin{eqnarray*}
v_{\mu} = \dfrac{\partial q}{\partial \mu} \Big \vert_{\mu = \mu_0, \sigma = \sigma_0} & = & q(x \vert \mu_0, \sigma^2_0)\dfrac{1}{2\sigma^2_0}(x-\mu_0) \\ 
v_{\sigma} = \dfrac{\partial q}{\partial \sigma} \Big \vert_{\mu = \mu_0, \sigma = \sigma_0} & = & q(x \vert \mu_0, \sigma_0)\left[ -\dfrac{1}{2}(\sigma_0)^{-1} + \dfrac{1}{2}\sigma_0^{-3}(x-\mu_0)^2 \right]\ \mbox{, where } -\infty < \sigma < \infty
\end{eqnarray*}
from which, we obtain an orthonormal basis of $T_{\theta_0}\Theta$,
\begin{eqnarray*}
w_{\mu} & = & q(x \vert \mu_0, \sigma^2_0) \dfrac{1}{\sqrt{\sigma^2_0}}(x-\mu_0) \\
w_{\sigma} & = &  q(x \vert \mu_0, \sigma^2_0) \dfrac{\sqrt{2}}{2} (\dfrac{(x-\mu_0)^2}{\sigma^2_0}-1)\\
\end{eqnarray*}
Finally, the negative gradient vector at $\theta_0$ is $\overrightarrow{v_0} = \dfrac{\langle w_{\mu}, \sqrt{p_0}\rangle w_{\mu}+\langle w_{\sigma}, \sqrt{p_0}\rangle w_{\sigma} }{\sqrt{1-\langle \theta_0, \sqrt{p_0}\rangle^2 }}$, where
\begin{eqnarray*}
\langle w_{\mu}, \sqrt{p_0}\rangle  & = &  \left(\dfrac{\pi}{2 \sigma_0^2} \right)^{-\frac{1}{4}} \dfrac{c_2}{\sqrt{\sigma_0^2}}\\
\langle w_{\sigma}, \sqrt{p_0}\rangle  & = &  \left(\dfrac{\pi}{2\sigma^2_0} \right)^{-\frac{1}{4}} (\dfrac{\sqrt{2}c_1}{2\sigma^2_0} - \dfrac{\sqrt{2}c_3}{2}) \\
\langle \theta_0, \sqrt{p_0}\rangle  & = &   \left(\dfrac{\pi}{2\sigma^2_0} \right)^{-\frac{1}{4}} c_3 
\end{eqnarray*}
Here, $c_1,c_2,c_3$ are integrals over $q_0$ and can be expressed as expectations with respect to $q(x \vert \mu_0, \sigma_0^2)$, 
\begin{eqnarray*}
c_1 & = &  E_{q_0^2} (\dfrac{(x-\mu_0)^2}{\sqrt{1+x^2}\exp(-\frac{1}{4\sigma^2_0}(x-\mu_0)^2)}) \\
 c_2 & = &  E_{q_0^2}  (\dfrac{x-\mu_0}{\sqrt{1+x^2}\exp(-\frac{1}{4\sigma^2_0}(x-\mu_0)^2)})  \\
 c_3 & = & E_{q_0^2} (\dfrac{1}{\sqrt{1+x^2}\exp(-\frac{1}{4\sigma^2_0}(x-\mu_0)^2)} \\
\end{eqnarray*}
We approximate these integrals using the Monte Carlo approximation method. 

\begin{figure}[t]
\begin{center}
\begin{tabular}{ccc}
{\includegraphics[width=0.3\textwidth]{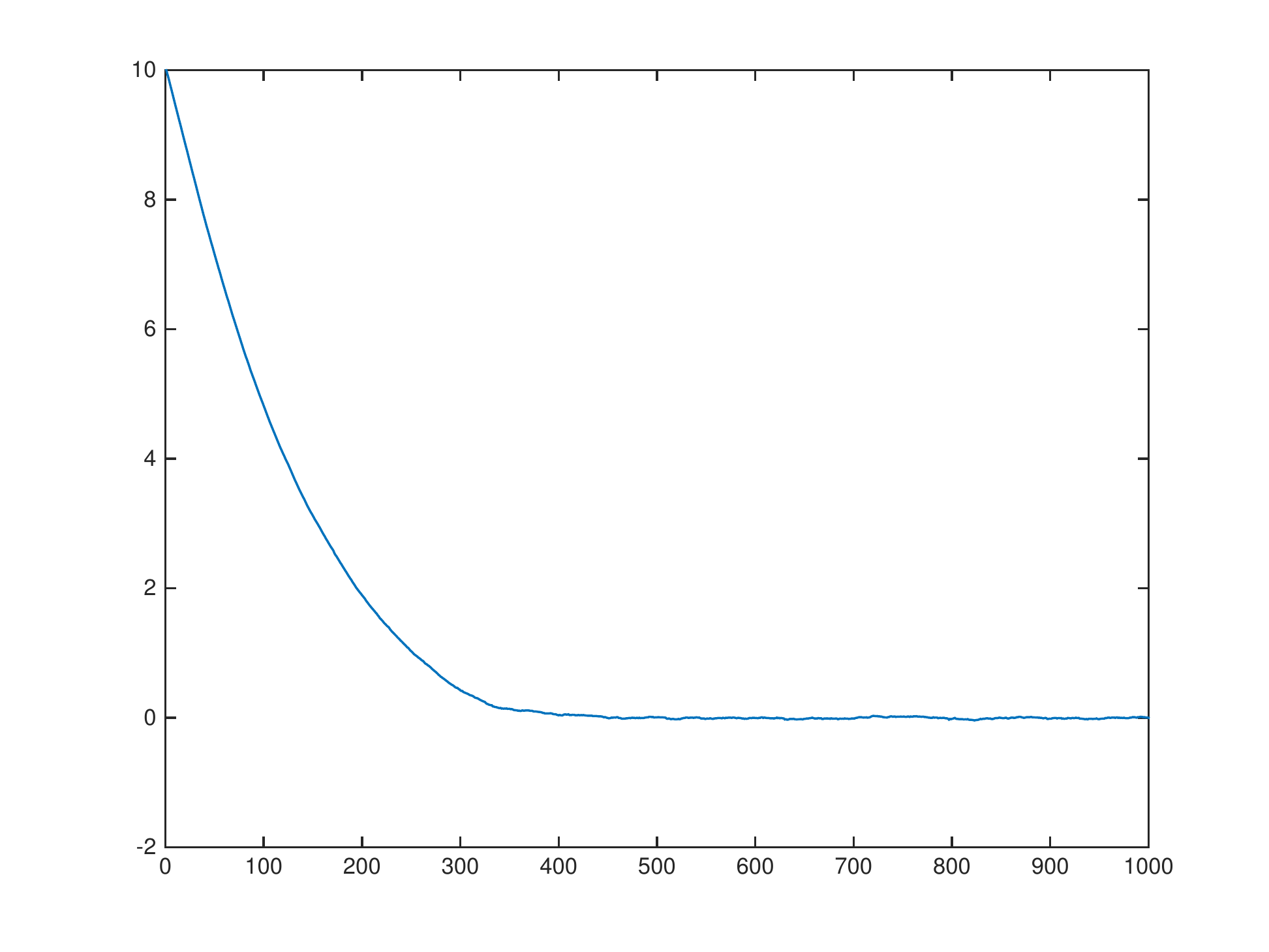}} &
{\includegraphics[width=0.3\textwidth]{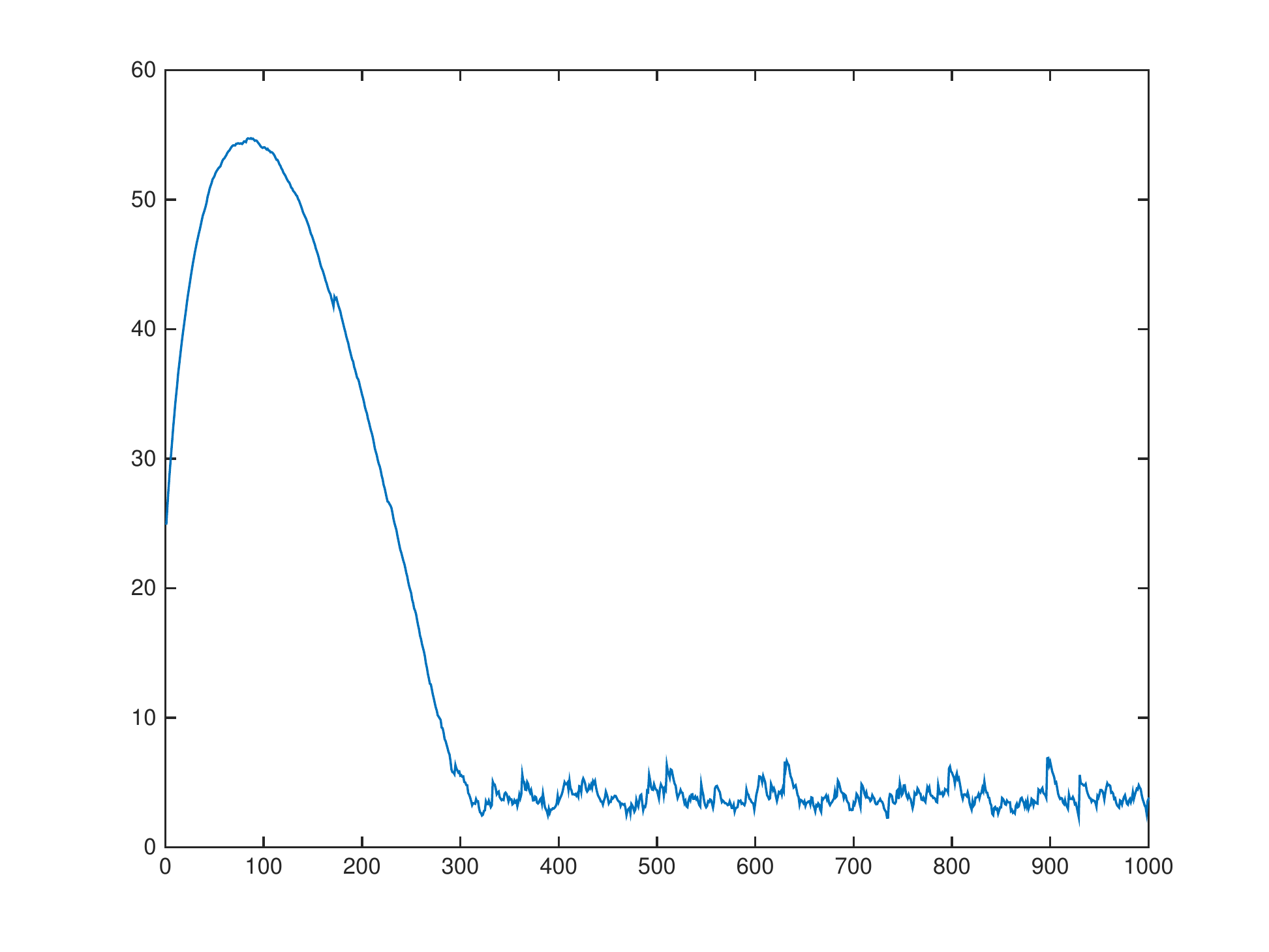}} &
{\includegraphics[width=0.3\textwidth]{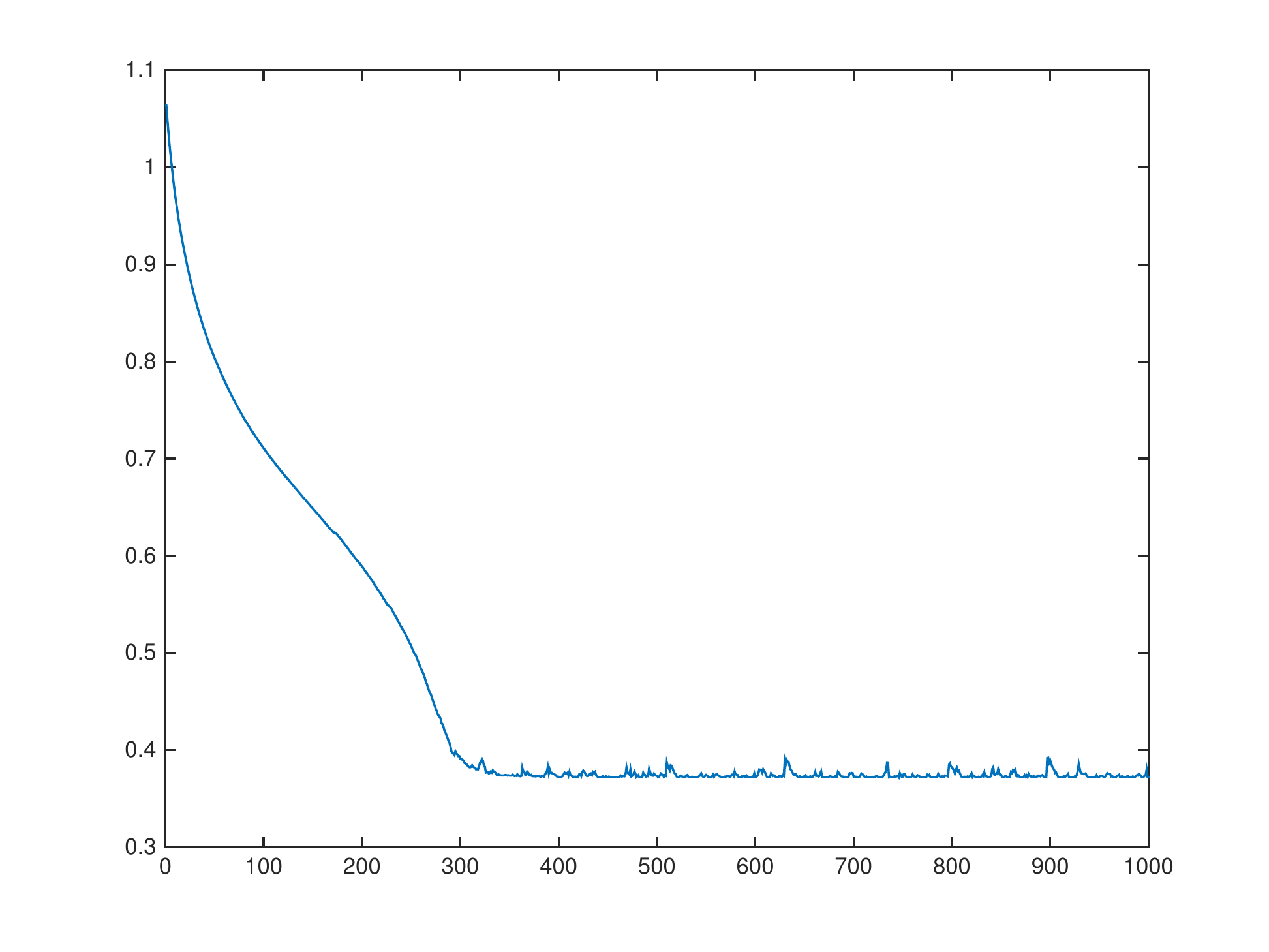}}\\
$\mu$ & $\sigma^2$ & $Distance$
\end{tabular}
\end{center}
\caption{Approximating $t(1)$ with $N(\mu, \sigma^2)$. As we can see, the distance reaches its minimum after 400 iterations, where $\mu$ and $\sigma^2$ converge to 0.0005 and 3.7468 respectively.}\label{fig:tDist}
\end{figure}

\begin{figure}[t]
\centering
\includegraphics[width=0.7\textwidth]{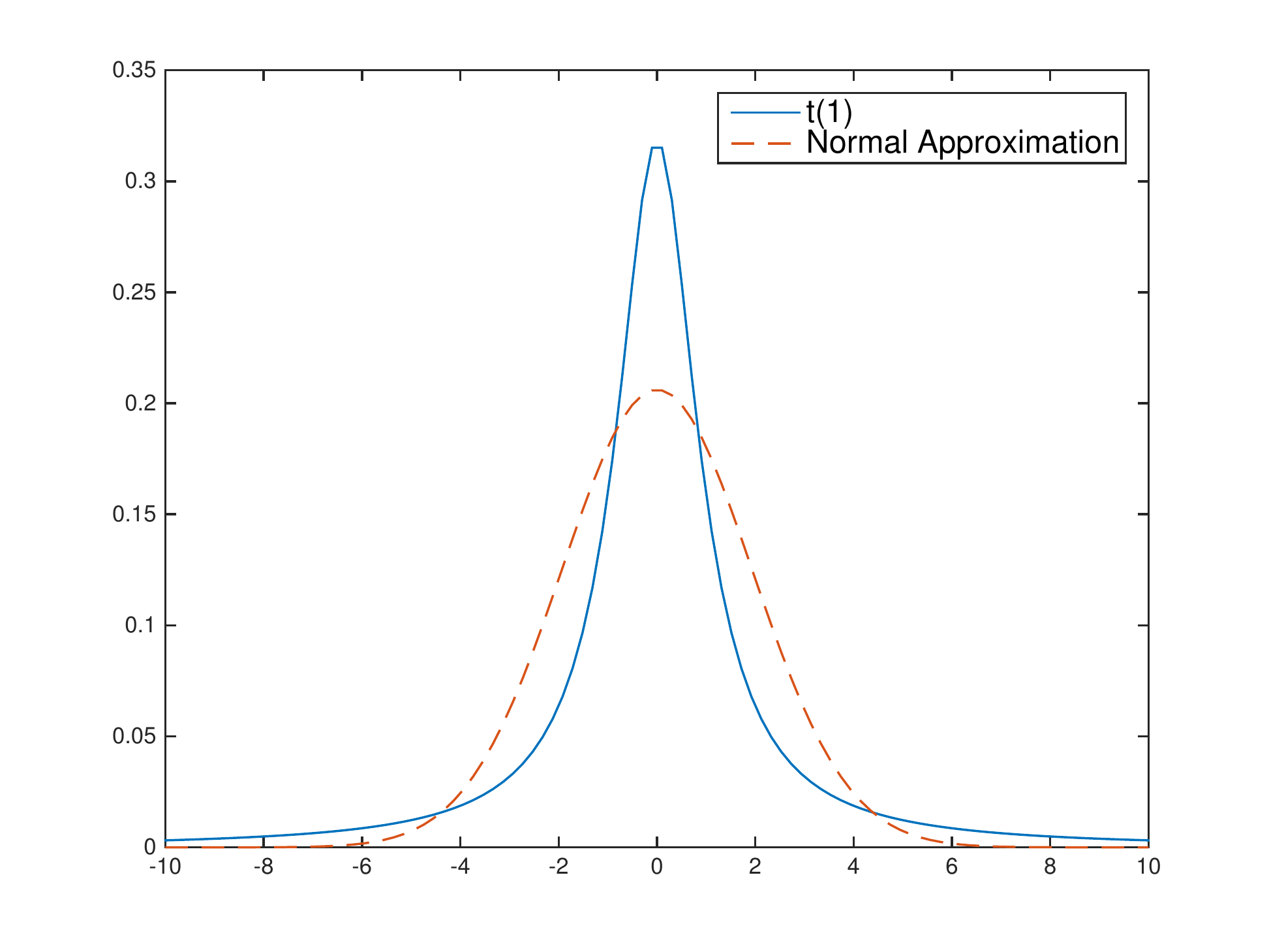}
\caption{Approximating $t(1)$ with $N(0.0005, 3.7468)$ based on our method.}\label{fig:tDistDensity}
\end{figure}

Ideally, we can update $\theta_0$ by following the geodesic flow $\gamma(t)$ with $\gamma(0) = \theta_0$ and $\gamma'(0) =  \overrightarrow{v_0}$. For simplicity, however, we follow an approximate path and update the parameters as follows:
\begin{eqnarray*}
\mu^{(t+1)} & = & \mu^{(t)}+\epsilon_{\alpha} \dfrac{\langle w_{\mu}, \sqrt{p_0}\rangle}{\sqrt{1-\langle \theta_0, \sqrt{p_0}\rangle^2 }} \\
{\sigma}^{(t+1)} & = & {\sigma}^{(t)}+\epsilon_{\beta} \dfrac{\langle w_{\sigma}, \sqrt{p_0}\rangle }{\sqrt{1-\langle \theta_0, \sqrt{p_0}\rangle^2 }} \\
{\sigma^2}^{(t+1)} & = & ({\sigma}^{(t+1)})^2
\end{eqnarray*}
%\begin{eqnarray*}
%\mu^{(t+1)} & = & \mu^{(t)}+\epsilon_{\alpha} \dfrac{c_2/\sqrt{\sigma_0^2}}{\sqrt{(\pi/2\sigma_0^2)^{\frac{1}{2}}-c_3^2}}\\
%{\sigma^2}^{(t+1)} & = & {\sigma^2}^{(t)}+\epsilon_{\beta} \dfrac{c_1/\sqrt{2}\sigma_0^2- c_3/\sqrt{2}}{\sqrt{(\pi/2\sigma_0^2)^{\frac{1}{2}}-c_3^2}}\\
%\end{eqnarray*}
See Appendix \ref{appd: tDist} for more details.

We initialize $(\mu, \sigma) = (10, 5)$ and set stepsizes $\epsilon_{\alpha} = 0.1, \epsilon_{\beta} = 5$. The sequence of parameters and the distance between the target and approximating distributions over 1000 iterations are shown in Figure \ref{fig:tDist}. The approximating distribution is converging to $N(0.0005, 3.7468)$ and the distance reaches its minimum after 400 iterations. Note that the stochastic path towards the end is due to the Monte Carlo approximation. The corresponding density functions for the target and approximating distributions are shown in Figure \ref{fig:tDistDensity}.

\subsection{Logistic Regression}
\begin{figure}[t]
\centering
\includegraphics[width=\textwidth]{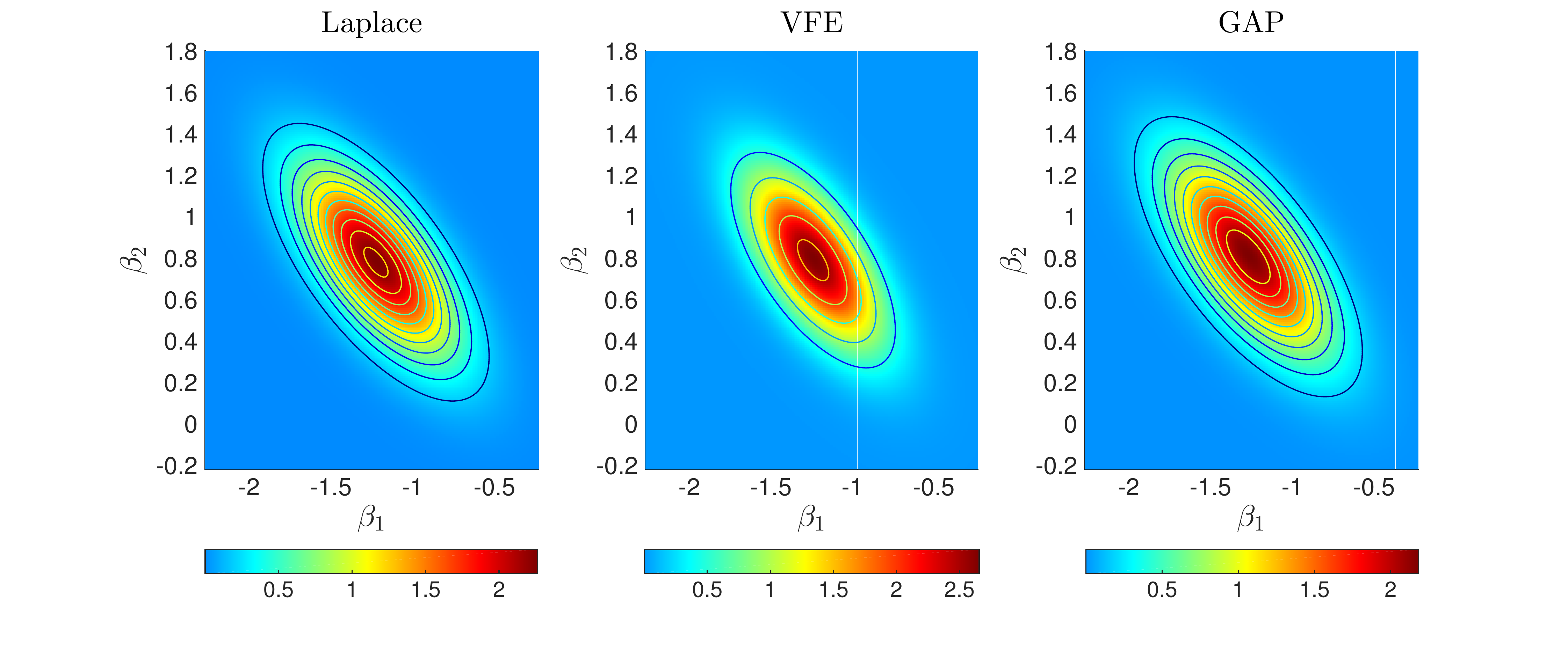}
\caption{Comparing our approximation method (GAP) to Laplace's approximation, and variational free energy (VFE) based on a logistic regression model.}\label{fig:comparison}
\end{figure}

For our next example, we consider Bayesian inference based on the following logistic regression model:
\begin{alignat*}{3}
& \mbox{Likelihood: } & y_i \vert X_i, \beta  \quad \sim & \quad \mbox{Bernoulli} (p_i = \dfrac{e^{X_i^T\beta}}{1+e^{X_i^T\beta}} )  \quad i = 1, \cdots, n\\
& \mbox{Prior: } & \beta  \quad  \sim  & \quad N_D(\mu_*, \Sigma_*) 
\end{alignat*}
The posterior distribution of model parameters and its square root are
\begin{eqnarray*}
p_0(\beta \vert X, y) & = & p(y \vert \beta) p(\beta)/p(y) \\
& = & \dfrac{1}{p(y)}\prod \limits_{i=1}^{n} \left( \dfrac{e^{X_i^T\beta}}{1+e^{X_i^T\beta}} \right) ^{y_i} \left(1- \dfrac{e^{X_i^T\beta}}{1+e^{X_i^T\beta}} \right) ^{1-y_i} (2\pi)^{-\frac{D}{2}} \vert \Sigma_* \vert^{-\frac{1}{2}}
\exp\left(-\frac{1}{2}( \beta - \mu_*)^T{\Sigma_*}^{-1}(\beta - \mu_*) \right) \\ 
\sqrt{p_0(\beta \vert X, y) } & = & p(y)^{-\frac{1}{2}}\left[ \prod \limits_{i=1}^{n} e^{\frac{y_i}{2} X_i^T\beta} (1+e^{X_i^T\beta})^{-\frac{1}{2}} \right]  (2\pi)^{-\frac{D}{4}} \vert \Sigma_* \vert^{-\frac{1}{4}}
\exp\left(-\frac{1}{4}( \beta - \mu_*)^T{\Sigma_*}^{-1}(\beta - \mu_*) \right) \\
\end{eqnarray*}
For approximation, we use the family of the $D$-dimensional Gaussian distributions and implement the algorithm presented in Section \ref{sec:algm1}. Figure \ref{fig:comparison} shows the normal approximation based on our method along with approximations obtained by Laplace's method and variational free energy (VFE) based on a dataset of size $N=100$ with $\beta_0 = 0.5, \beta_1 = -1.5$, and $\beta_2 = 1$. For this example, we generated $(x_1, x_2)$ from a bivariate normal distribution with zero means, unit variances and correlation 0.7.
%$N \left( \begin{pmatrix}
%0 \\ 0
%\end{pmatrix}, 
%\begin{pmatrix}
%1&0.7 \\0.7 &1
%\end{pmatrix}
%\right)$. 

As expected, the approximating distribution based on variational free energy is more compact than the true distribution \cite{mackay03}. Note that here we used a local variational method, where a lower bound is found for a part of the entire probabilistic model to simplify the approximation (\cite{Bishop2007} \cite{Jaakkola2000}). For Bayesian logistic regression, a lower bound $h(\beta, \xi)$ for $p(y \vert \beta)$ can be derived using the convex duality framework, where $\xi$ are variational parameters. The variational posterior then can be obtained by maximizing $ \mathcal{L}(\xi) \equiv \ln \int h(\beta, \xi)p(\beta) d \beta \leqslant  \ln \int p(y \beta) p(\beta) d\beta = \ln p (y)$ using the Expectation-Maximization (EM) algorithm.

%$ h(\beta,\xi)$ takes the form of exponential quadratic function and $\xi$ is a collection of local variational parameter $\xi_i$  corresponding to each data point $\beta^Tx_i$. We then approximate the log evidence by $$\ln p (y) = \ln \int p(y \beta) p(\beta) d\beta \geqslant \ln \int h(\beta, \xi)p(\beta) d \beta \equiv \mathcal{L}(\xi)$$. 

For this example, our approximating distribution is almost the same as what we obtain from Laplace's method. However, as illustrated by our next example, this is not the case in general. 
 
\subsection{Approximating a bimodal distribution}
\begin{figure*}[t]
\begin{center}
\begin{tabular}{cc}
\includegraphics[width=0.5\textwidth]{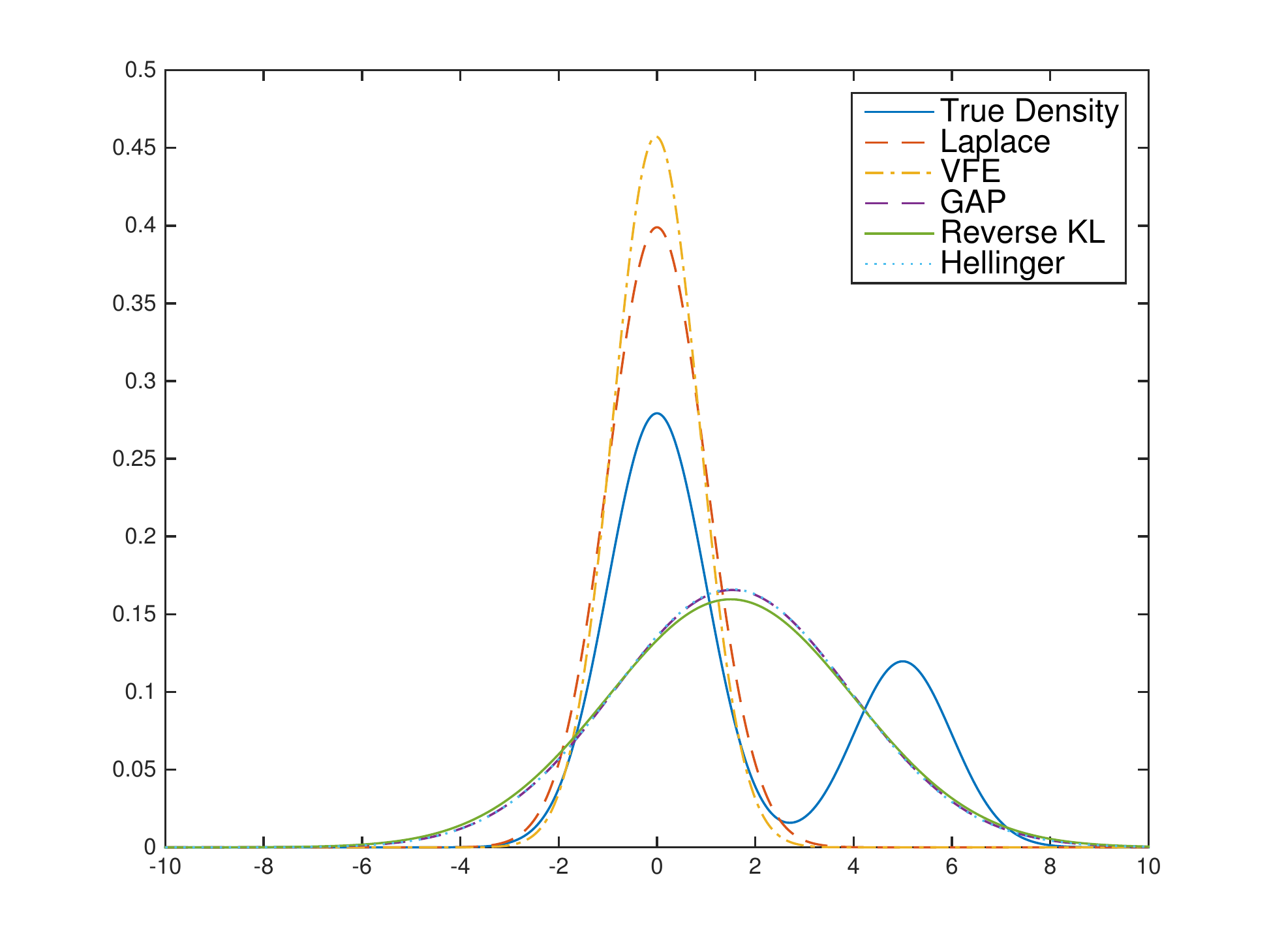} & \includegraphics[width=0.5\textwidth]{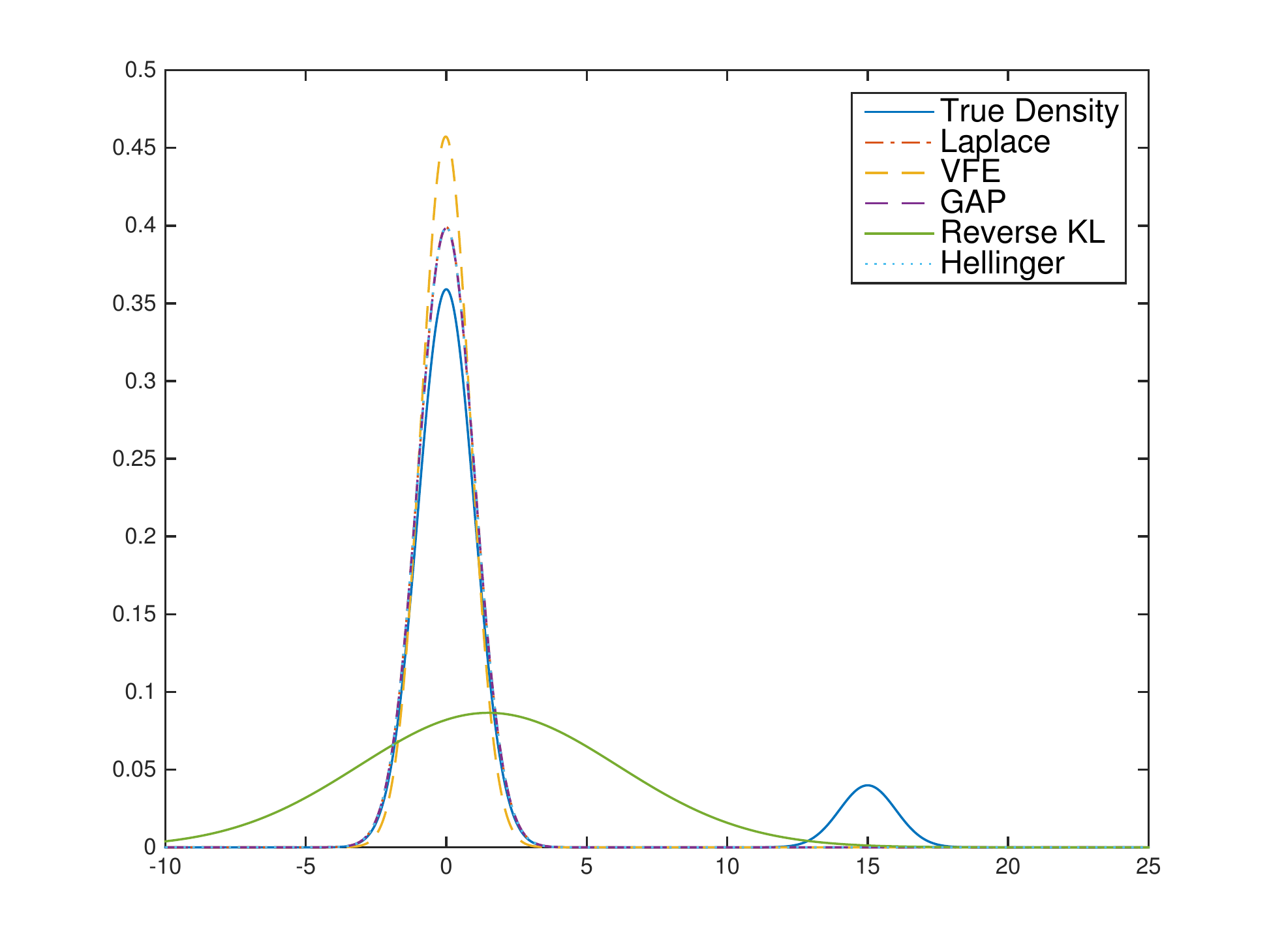}\\
(a) $x \sim 0.7N(0,1)+0.3N(5,1)$ & (b) $x \sim 0.9N(0,1)+0.1N(15,1)$
\end{tabular}
\end{center}
\caption{Approximating bimodal distributions using our method (GAP),  Laplace's approximation, variational free energy, Reverse KL and the Hellinger distance.}\label{fig:bimodal}
\end{figure*}

For our final example, we use our method to find a univariate Gaussian approximation to mixture of normals. First, we use our method to find a normal approximation to the following bimodal distribution: $$x \sim 0.7N(0,1)+0.3N(5,1).$$ The left panel of Figure \ref{fig:bimodal} compares the result of our model to those based on Laplace's approximation and $\alpha$-divergence, for different values of $\alpha$ (KL-divergence, reverse KL-divergence, and the Hellinger distance). As we can see, while Laplace's approximation and variational free energy (VFE) capture the first mode only (hence, underestimating the variance), our method increases the variance to cover both modes. Similar results are obtained based on reverse KL-divergence and the Hellinger distance. As expected, the results based on GAP and the Hellinger distance are almost indistinguishable. 
  
Recall that $\lim_{\alpha \rightarrow 0} D_{\alpha} (p \Vert p') = KL(p' \Vert p)$ and $\lim_{\alpha \rightarrow 1} D_{\alpha} (p \Vert p') = KL(p \Vert p')$. When $\alpha < 0$, minimizing $D_{\alpha} (p \Vert p')$ tends to give zero-forcing results, because when $p$ is close to zero, $p'$ also has to be close to zero to avoid large penalties. Therefore, the VFE method in this case captures a single mode. However, when $\alpha >1$, the result is zero-avoiding, i.e., $p'$ tends to be greater than zero in regions where $p$ is greater than zero. Thus, results based on reverse KL will average across both modes. When $0 < \alpha <1$, the results are in between: they are neither zero-forcing nor zero-avoiding, so it tends to cover across modes but will fail to find modes that are far from the main mass  \cite{Minka2005}. To see this, we consider another mixture distribution which has a mode far from the main mass: $$x \sim 0.9N(0,1)+0.1N(15,1).$$ The right panel of Figure \ref{fig:bimodal} shows the corresponding results. Here, we observe that the Hellinger distance fails to capture the far mode, as opposed to reverse KL. As discussed before, minimizing spherical Fisher distance is equivalent to minimizing the Hellinger distance, so the approximating distribution based of our method (GAP) is similar to the distribution based on the the Hellinger distance in both examples.

%In addition, for minimizing reverse KL, when the proposed model is restricted to be exponential family member, the result is moment matching. It is interesting to observe that our method as well as minimizing the Hellinger distance also display the property of moment matching. 

\section{Discussion}
\label{sec:discussion}
We have proposed a novel framework for approximating posterior distributions and illustrated its performance using several examples. Application of our method, however, can go well beyond what discussed here. As a deterministic approximation approach, our method has the potential to scale better compared to MCMC methods. Compared to other deterministic approaches, our method's flexibility and generalizability could lead to substantially more accurate approximation of posterior distribution, which in turn would lead to more accurate statistical inference.  
 
Although in this paper we limited the class of approximating distributions to normals, our method can be generalized to other approximating distributions as long as we could obtain the orthonormal basis $\{ w_i \}$ with respect to each particular point $\theta$ on $\Theta$. For example, we can set $\Theta$ to be mixture of normals. This would allow for more flexibility in approximating target distributions.  

To make our method more practical, we should substantially improve its computational efficiency. Currently, the computational cost of our method is mainly dominated by finding the orthonormal basis of $T_{\theta}\Theta$. Also, finding alternatives to Monte Carlo method for approximating intractable integrals in our algorithm could help to make our method more efficient.

Finally, we need to further study the properties of our proposed method and its connection to other approximation approaches such as those based on $\alpha$-divergence. It will also be of great importance to identify classes of approximating distributions that lead to convex optimization problems. For non-convex problems, we need to improve our numerical optimization method to avoid falling into local minima. 

%For the non-convex problems, randomized methods could be incorporated to avoid falling into local minimums.
  
%Secondly, in the process of calculating parameter updates, we need to evaluate intractable integrals. In our implementation, Monte Carlo methods are used. However, ... will be more appreciated.
%
%In addition, even though our method doesn't have to obtain posterior samples as MCMC does, the computational cost for now is still expensive. Other than the reason of using Monte Carlo method to obtain intractable integrals, another reason is that the matrix expression of Gram–Schmidt process is employed to obtain the orthonormal basis. The complexity is $O(D^4)$ for each iteration, i.e. $O(D^4) \times$ the number of iterations it takes to converge. Better algorithm should be developed to improve this aspect.

\newpage
\appendix
\begin{center}
{\Huge Appendix}
\end{center}

%\section{Matrix operations}
%\begin{eqnarray*}
%\dfrac{\partial det(X)}{\partial X} & = & det(X)(X^{-1})^T \\
%\dfrac{\partial a^TX^{-1}b}{\partial X} & = & -X^{-T}ab^TX^{-T}\\
%\dfrac{\partial (X^{-1})_{kl}}{\partial X_{ij}} & = & -(X^{-1})_{ki}(X^{-1})_{jl} \\
%\dfrac{\partial (X^TBX)_{kl}}{\partial X_{ij}} & = & \delta_{lj}(X^TB)_{ki}+\delta_{kj}(BX)_{il} \\
%\end{eqnarray*}

\section{An illustrative example with analytical solution}\label{illustExample1}
We now provide the details for the illustrative example with analytical solution discussed in Section \ref{VB-AFG}. For this problem we have,
\begin{alignat*}{3}
& \mbox{Posterior density: } & p &  \propto   P(x \vert \mu, \tau) P(\mu \vert \tau) P(\tau) \\
& & & = (2\pi)^{-\frac{n}{2}}\tau^{\frac{n}{2}}\exp\left(-\dfrac{\tau}{2} \sum (x_i-\mu)^2 \right) \cdot (2\pi)^{-\frac{1}{2}}(\lambda_0\tau)^{\frac{1}{2}}\exp(-\dfrac{\lambda_0\tau}{2}(\mu-\mu_0)^2) \\
& & &  \dfrac{\beta_0^{\alpha_0}}{\Gamma(\alpha_0)}\tau^{\alpha_0-1}\exp(-\beta_0\tau) \\
& \mbox{Its square root: } & \sqrt{p} & \propto  (2\pi)^{-\frac{n}{4}}\tau^{\frac{n}{4}}\exp\left(-\dfrac{\tau}{4} \sum (x_i-\mu)^2 \right) \cdot (2\pi)^{-\frac{1}{4}}(\lambda_0\tau)^{\frac{1}{4}}\exp(-\dfrac{\lambda_0\tau}{4}(\mu-\mu_0)^2) \\
& & &  \sqrt{\dfrac{\beta_0^{\alpha_0}}{\Gamma(\alpha_0)}}\tau^{\frac{\alpha_0-1}{2}}\exp(-\frac{\beta_0\tau}{2}) \\
& \mbox{Approximating density: } & p' & =  P'(\mu \vert \tau)P'(\tau) \\
& & & = (2\pi)^{-\frac{1}{2}}(\lambda_N\tau)^{\frac{1}{2}}\exp(-\dfrac{\lambda_N\tau}{2}(\mu-\mu_N)^2)  \dfrac{\beta_N^{\alpha_N}}{\Gamma(\alpha_N)}\tau^{\alpha_N-1}\exp(-\beta_N\tau) \\
& \mbox{Its square root: } & \sqrt{p'}  & =  (2\pi)^{-\frac{1}{4}}(\lambda_N\tau)^{\frac{1}{4}}\exp(-\dfrac{\lambda_N\tau}{4}(\mu-\mu_N)^2)  \sqrt{\dfrac{\beta_N^{\alpha_N}}{\Gamma(\alpha_N)}}\tau^{\frac{\alpha_N-1}{2}}\exp(-\frac{\beta_N\tau}{2}) 
\end{alignat*}

The spherical Fisher distance between $p$ and $p'$ is
\begin{eqnarray*}
d_{SF}(p,p') & = & \arccos \int \int \sqrt{p p'} d\mu d\tau  \\
& \propto & \arccos \int \int f(\mu, \tau)  \sqrt{\dfrac{\beta_N^{\alpha_N}}{\Gamma(\alpha_N)}}  \lambda_N^{\frac{1}{4}}\dfrac{\Gamma(\alpha^*)\sqrt{2\pi}}{{\beta^*}^{\alpha^*} {\lambda^*}^{\frac{1}{2}}} d\mu d\tau \\
\mbox{where } f(\mu, \tau) & = & \dfrac{{\beta^*}^{\alpha^*} {\lambda^*}^{\frac{1}{2}}}{\Gamma(\alpha^*)\sqrt{2\pi}}\tau^{\alpha^*-\frac{1}{2}}\exp(-\beta^*\tau) \exp(-\dfrac{\lambda^*\tau(\mu-\mu^*)^2}{2}) \\
& & \mbox{ is the joint Normal-Gamma density of } (\mu, \tau) \mbox{ parameterized by } (\mu^*, \lambda^*, \alpha^*, \beta^*): \\
\mu^* & = & \dfrac{n\bar{x}+\lambda_0\mu_0+\lambda_N\mu_N}{n+\lambda_0+\lambda_N} \\
\lambda^* & = & \dfrac{n+\lambda_0+\lambda_N}{2} \\
 \alpha^* & = &  \dfrac{n}{4}+\dfrac{\alpha_0+\alpha_N}{2}\\
\beta^* & = &  \dfrac{S}{4}+\dfrac{\beta_0+\beta_N}{2} + \dfrac{n\lambda_0(\bar{x}-\mu_0)^2+n\lambda_N(\bar{x}-\mu_N)^2+\lambda_0\lambda_N(\mu_0-\mu_N)^2}{4(n+\lambda_0+\lambda_N)},  S   =  \sum (x_i-\bar{x})^2 
\end{eqnarray*}
 
Therefore, we have: $d_{SF}(p, p')  \propto \arccos  \sqrt{\dfrac{\beta_N^{\alpha_N}}{\Gamma(\alpha_N)}}\dfrac{\lambda_N^{\frac{1}{4}} \Gamma(\alpha^*)}{{\beta^*}^{\alpha^*}{\lambda^*}^{\frac{1}{2}}}  \equiv  \arccos g( \mu_N, \lambda_N, \alpha_N, \beta_N) $.

Since $arccos$ function is monotone decreasing and $log$ function is monotone increasing, minimizing the spherical Fisher distance between $p$ and $p'$ with respect to $ \mu_N,  \lambda_N, \alpha_N, \beta_N$ is equivalent to maximizing:
$$ \log  g( \mu_N, \lambda_N,\alpha_N, \beta_N)   = \dfrac{\alpha_N}{2}\log \beta_N-\dfrac{1}{2}\log \Gamma (\alpha_N)+\dfrac{1}{4} \log \lambda_N + \log \Gamma (\alpha^*) - \alpha^* \log \beta^* -\dfrac{1}{2} \log \lambda^* . $$

To maximize the above function, we need to solve $\dfrac{\partial \log g}{\partial \mu_N} = 0 $, $\dfrac{\partial \log g}{\partial \lambda_N} = 0 $, $\dfrac{\partial \log g}{\partial \alpha_N} = 0 $, $\dfrac{\partial \log g}{\partial \beta_N} = 0 $. Note that
\begin{eqnarray*}
\dfrac{\partial \alpha*}{\partial \alpha_N} & = & \dfrac{1}{2}, \quad  \dfrac{\partial \beta*}{\partial \beta_N} = \dfrac{1}{2}, \quad  \dfrac{\partial \lambda^*}{\partial \lambda_N} = \dfrac{1}{2}\\
\dfrac{\partial \beta*}{\partial \lambda_N} &  = &  \dfrac{\left(n(\mu_N-\bar{x}) + \lambda_0 (\mu_N-\mu_0) \right)^2}{4(n+\lambda_0+\lambda_N)^2}, \quad
\dfrac{\partial \beta*}{\partial \mu_N} = \dfrac{\lambda_N\mu_N(n+\lambda_0)-\lambda_N(n\bar{x}+\lambda_0\mu_0)}{2(n+\lambda_0+\lambda_N)}
\end{eqnarray*}
To find the optimal $\mu_N, \lambda_N$, we solve
\begin{eqnarray*}
\dfrac{\partial \log g}{\partial \mu_N}  =  0    & \Rightarrow  &
 \mu_N  =  \dfrac{n\bar{x}+\lambda_0\mu_0}{n+\lambda_0} \\
\dfrac{\partial \log g}{\partial \lambda_N} = 0   & \Rightarrow & \dfrac{1}{4 \lambda_N}-\dfrac{1}{4 \lambda^*}  -\dfrac{\alpha^*}{\beta^*} \dfrac{\left(n(\mu_N-\bar{x}) + \lambda_0 (\mu_N-\mu_0) \right)^2}{4(n+\lambda_0+\lambda_N)^2} = 0\\
\mbox{But given } \mu_N  =  \dfrac{n\bar{x}+\lambda_0\mu_0}{n+\lambda_0} \mbox{, we have } \lambda_N  = \lambda^*   & \Rightarrow  & \lambda_N =  \lambda_0+ n  
\end{eqnarray*}
Finally, we find  the optimal $\alpha_N$ and $\beta_N$ as follows:
\begin{eqnarray*}
\dfrac{\partial \log g}{\partial \alpha_N} =  0 & \Rightarrow & \log \beta_N - \log \beta^* = \psi (\alpha_N) - \psi (\alpha^*),   \mbox{ where } \psi(x) =\frac{d}{dx} \ln{\Gamma(x)}= \frac{\Gamma'(x)}{\Gamma(x)} \\
\dfrac{\partial \log g}{\partial \beta_N} = 0 & \Rightarrow &  \dfrac{\beta_N}{\beta^*} - \dfrac{\alpha_N}{\alpha^*} = 0 
\end{eqnarray*}
Trivially, we solve $ \log \alpha_N - \log \alpha^*  =   \psi (\alpha_N) - \psi (\alpha^*) $ by setting $\alpha_N = \alpha^*  $, so the optimal $\alpha_N  = \alpha_0+\dfrac{n}{2} $. Finally, we have:
 \begin{eqnarray*}
 \beta_N & = & \beta^* \\
 & = & \dfrac{ \alpha_N}{(n+\lambda_0+\lambda_N)(n+2\alpha_0)}  \bigg( (n+\lambda_0+\lambda_N)(S+2\beta_0)  + n\lambda_0(\bar{x}-\mu_0)^2 +n\lambda_N(\bar{x}-\mu_N)^2+\lambda_0\lambda_N(\mu_0-\mu_N)^2 \bigg ) \\
 & = & \beta_0 + \dfrac{S}{2} + \dfrac{n\lambda_0(\bar{x}-\mu_0)^2+n\lambda_N(\bar{x}-\mu_N)^2+\lambda_0\lambda_N(\mu_0-\mu_N)^2}{2(n+\lambda_0+\lambda_N)}\\
& = & \beta_0 + \dfrac{S}{2} + \dfrac{n\lambda_0(\bar{x}-\mu_0)^2}{2(n+\lambda_0)} 
\end{eqnarray*}

\section{Gaussian approximation}
\label{appd: algm1}
For our general Gaussian approximation methods, the algorithm involves several steps as described below. These steps are summarized in Algorithm~\ref{alg:innerMu} and Algorithm~\ref{alg:innerSigma2}.   

\paragraph{Finding non-orthonormal basis of $T_{\theta_0}\Theta$}
In this paper, we define the  derivatives of the map $f: \mathbb{R}^n \rightarrow \mathbb{R}^m $ as $ [\frac{d\mathbf{f} }{d \mathbf{x}}]_{ij} = \frac{\partial f_i(\mathbf{x})}{x_j}$ and we use the following notations \cite{fackler2005notes}, \cite{petersen2008matrix}:

\begin{minipage}[t]{0.9\textwidth}
\begin{flalign*}
& A \otimes B: \mbox{Kronecker product} &\\
& A \circ B: \mbox{ Hadamard (elementwise) product}  & \\
& vec:  \mbox{an operator which vectorizes the matrix column-wisely   } & \\
& vech: \mbox{ an operator which vectorizes the matrix column-wisely but excludes the upper part of the matrix} & \\
&T_{m,n}: T_{m,n} vec(A_{m \times n}) = vec(A^T) & \\
&R_n: R_n vec(A_{n \times n}) = vec(A_{n \times n} \circ I) & \\
& S_n:  vech(A_{n \times n}) = S_n vec(A_{n \times n}) 
\end{flalign*}
\end{minipage} 
\vspace{12pt}

\noindent
We calculate the basis as follows: 
\begin{eqnarray*}
 v_{\mu}  =   \dfrac{\partial q}{\partial \mu} \Big \vert_{\mu = \mu_0, L = L_0} & = & q(z \vert \mu_0, \Sigma_0)\dfrac{1}{2} (z-\mu_0)^T \Sigma_0^{-1}, \quad ({1 \times D} \mbox{ vector}) \\
v_l  =   \dfrac{\partial q}{\partial vech(L)} \Big \vert_{\mu = \mu_0, L = L_0} & = &   q(z \vert \mu_0, \Sigma_0)  \left[ -\dfrac{1}{4} vec(\Sigma_0^{-T})^{T} +\dfrac{1}{4}((z-\mu_0)^T \otimes (z-\mu_0)^T)(\Sigma_0^{-T} \otimes \Sigma_0^{-1})\right] \\
& & \left[I+T_{D,D}^T-R_D^T \right] \left[ (I_D \otimes L_0) T_{D,D} + (L_0 \otimes I_D) \right] S_D^T , \quad  (1 \times {\dfrac{D(D+1)}{2}} \mbox{ vector}) 
\end{eqnarray*}
The basis $v_l$ is obtained by using the chain rule:
$$ \dfrac{\partial q}{\partial vech(L)}  =  \dfrac{\partial q}{\partial  vec(\Sigma)} \dfrac{d vec(\Sigma)}{d vech(L)} $$
If we assume that $\Sigma$ is an unstructured matrix (i.e. the entries of $\Sigma$ are entirely independent), we have:
\begin{eqnarray*}
\dfrac{d q}{d vec(\Sigma)} & = &  q(z \vert \mu, \Sigma)  \left[ -\dfrac{1}{4} vec(\Sigma^{-T})^{T} +\dfrac{1}{4}((z-\mu)^T \otimes (z-\mu)^T)(\Sigma^{-T} \otimes \Sigma^{-1})\right] 
\end{eqnarray*}
But because $\Sigma$ is symmetric, the general rules do not apply. Instead, we have  \footnote{We use symbol $d$ for derivatives with respect to unstructured $\Sigma$ and symbol $\partial$ for symmetric $\Sigma$} $$\dfrac{\partial q}{\partial \Sigma} =  \dfrac{d q}{d \Sigma} + \left( \dfrac{d q}{d \Sigma} \right)^T -   \dfrac{d q}{d \Sigma} \circ  I$$ 
Correspondingly,
\begin{eqnarray*}
\dfrac{\partial q}{\partial vec(\Sigma)}  & = & vec\left( \dfrac{d q}{d \Sigma} \right)^T + vec\left( \dfrac{d q}{d \Sigma}^T \right)^T - vec\left(\dfrac{d q}{d \Sigma} \circ  I \right)^T \\
& = & \dfrac{d q}{d vec(\Sigma)}  +  \dfrac{d q}{d vec(\Sigma)} T_{D,D} ^T - \dfrac{d q}{d vec(\Sigma)}  R_D^T 
\end{eqnarray*} 
Finally, we have: 
\begin{eqnarray*}
\dfrac{d vec(\Sigma)}{d vech(L)} &  = &  \dfrac{d vec(LL^T)}{d vech(L)} \\
& = & \left[ (I_D \otimes L) T_{D,D} + (L \otimes I_D) \right]  \dfrac{d vec(L)}{d vech(L)}  \\
& = & \left[ (I_D \otimes L) T_{D,D} + (L \otimes I_D) \right] S_D^T 
\end{eqnarray*} 
 
\paragraph{Finding the inner products of the basis}
Notice that the inner product in $\mathcal{Q}$ is defined as: $$\langle  \varphi, \phi \rangle = \int^{+\infty}_{-\infty} \varphi \phi dz$$ 
We can then find the corresponding inner products,
\begin{eqnarray*}
\langle v_{\mu_i}, v_{\mu_j} \rangle & = & \dfrac{1}{4} (\Sigma_0^{-1})_{ij} =  \dfrac{1}{4} \sigma^{ij}\\
\langle v_{\mu_i},{v_l}_j \rangle & = &  0 \\
\langle {v_l}_i, {v_l}_j \rangle 
& = &  B_{ij}
\end{eqnarray*}
where $B$ is the inner product matrix for $v_l$. We show how to obtain $B$. For simplicity, we denote $v_l  =  q(z \vert \mu_0, \Sigma_0) W_D U_D V_D$, where
\begin{eqnarray*}
W_D & = &    -\dfrac{1}{4} vec(\Sigma_0^{-T})^{T} +\dfrac{1}{4}((z-\mu_0)^T \otimes (z-\mu_0)^T)(\Sigma_0^{-T} \otimes \Sigma_0^{-1}) \\
U_D & = & I+T_{D,D}^T-R_D^T  \\
V_D & = & \left[ (I_D \otimes L_0) T_{D,D} + (L_0 \otimes I_D) \right] S_D^T  
\end{eqnarray*}
Thus, $B = E_{q^2(z \vert \mu_0, \Sigma_0)} (V_D^TU_D^TW_D^TW_DU_DV_D) =  V_D^TU_D^TE_{q^2(z \vert \mu_0, \Sigma_0)}(W_D^TW_D)U_DV_D  $. We can calculate $E (W_D^TW_D) $ as follows:
 \begin{eqnarray*} 
E (W_D^TW_D) & = & E( \left[ -\dfrac{1}{4} vec(\Sigma_0^{-T})  +\dfrac{1}{4}(\Sigma_0^{-T} \otimes \Sigma_0^{-1}) ((z-\mu_0) \otimes (z-\mu_0)) \right] \\
& & \left[ -\dfrac{1}{4} vec(\Sigma_0^{-T})^{T} +\dfrac{1}{4}((z-\mu_0)^T \otimes (z-\mu_0)^T)(\Sigma_0^{-T} \otimes \Sigma_0^{-1})\right]  ) \\
& = & \dfrac{1}{16} vec(\Sigma_0^{-1}) vec(\Sigma_0^{-1})^{T} -  \dfrac{1}{16} vec(\Sigma_0^{-1}) E ((z-\mu_0)^T \otimes (z-\mu_0)^T)(\Sigma_0^{-1} \otimes \Sigma_0^{-1}) \\
 & & -\dfrac{1}{16} (\Sigma_0^{-T} \otimes \Sigma_0^{-1}) E ((z-\mu_0) \otimes (z-\mu_0)) vec(\Sigma_0^{-1})^{T} \\
 & & + \dfrac{1}{16}(\Sigma_0^{-1} \otimes \Sigma_0^{-1}) E((z-\mu_0) \otimes (z-\mu_0)) ((z-\mu_0)^T \otimes (z-\mu_0)^T)(\Sigma_0^{-1} \otimes \Sigma_0^{-1}) 
\end{eqnarray*} 
Notice that we assume $z$ follows a normal distribution, then according to Isserlis' theorem, we have:
\begin{eqnarray*}
E\left( (z_i - \mu_i)(z_j-\mu_j)(z_k-\mu_k) \right) & = &  0\\
E\left( (z_i - \mu_i)(z_j-\mu_j)(z_k-\mu_k)(z_l-\mu_l) \right) & = & \sigma_{ij}\sigma_{kl} +\sigma_{ik}\sigma_{jl} +\sigma_{jk}\sigma_{il} 
\end{eqnarray*}
Therefore,
 \begin{eqnarray*}
 E((z-\mu)^T \otimes (z-\mu)^T) & = & vec (\Sigma^T)^T\\
  E((z-\mu) \otimes (z-\mu) ) & = & vec (\Sigma^T)\\
  E((z-\mu)(z-\mu)^T \otimes (z-\mu) (z-\mu)^T) & = & \left[ (\Sigma \otimes \Sigma)_{ijkl} \right] +  \left[ (\Sigma \otimes \Sigma)_{ikjl} \right] +  \left[ (\Sigma \otimes \Sigma)_{iljk} \right] 
\end{eqnarray*}
$\left[ (\Sigma \otimes \Sigma)_{ijkl} \right], \left[ (\Sigma \otimes \Sigma)_{ikjl} \right], \left[ (\Sigma \otimes \Sigma)_{iljk} \right] $ correspond to different permutations of $\Sigma \otimes \Sigma$ such that the matrix elements are $\sigma_{ij}\sigma_{kl}, \sigma_{ik}\sigma_{jl}, \sigma_{jk}\sigma_{il}$ respectively. Finally, we have
\begin{eqnarray*}
 E (W_D^TW_D) & = & \dfrac{1}{16} vec(\Sigma_0^{-1}) vec(\Sigma_0^{-1})^{T} -  \dfrac{1}{16} vec(\Sigma_0^{-1})  vec (\Sigma_0)^T (\Sigma_0^{-1} \otimes \Sigma_0^{-1}) -\dfrac{1}{16} (\Sigma_0^{-T} \otimes \Sigma_0^{-1}) vec (\Sigma_0) vec(\Sigma_0^{-1})^{T} \\
  & & + \dfrac{1}{16}(\Sigma_0^{-1} \otimes \Sigma_0^{-1})  (\left[ (\Sigma_0 \otimes \Sigma_0)_{ijkl} \right] +  \left[ (\Sigma_0 \otimes \Sigma_0)_{ikjl} \right] +  \left[ (\Sigma_0 \otimes \Sigma_0)_{iljk} \right] ) (\Sigma_0^{-1} \otimes \Sigma_0^{-1}) 
\end{eqnarray*}
\paragraph{Orthonormalizing the basis}
Because $\langle v_{\mu_i},{v_l}_j \rangle  = 0$, we only need to orthonormalize the set of basis $\{ v_{\mu_i} \}_{i=1}^D$ and $\{{v_l}_i \}_{i=1}^{\frac{D(D+1)}{2}}$ respectively. We use the Gram-Schmidt process to find the orthonormal basis. For $\{v_{\mu_i} \}_{i=1}^D$, for example, we have
\begin{eqnarray*}
 w_{\mu_j} & = & \frac{1}{\sqrt{D_{j-1} D_j}} \begin{vmatrix}
\langle \mathbf{v}_{\mu_1}, \mathbf{v}_{\mu_1} \rangle & \langle \mathbf{v}_{\mu_2}, \mathbf{v}_{\mu_1} \rangle & \dots & \langle \mathbf{v}_{\mu_j}, \mathbf{v}_{\mu_1} \rangle \\
\langle \mathbf{v}_{\mu_1}, \mathbf{v}_{\mu_2} \rangle & \langle \mathbf{v}_{\mu_2}, \mathbf{v}_{\mu_2} \rangle & \dots & \langle \mathbf{v}_{\mu_j}, \mathbf{v}_{\mu_2} \rangle \\
\vdots & \vdots & \ddots & \vdots \\
\langle \mathbf{v}_{\mu_1}, \mathbf{v}_{\mu_{j-1}}  \rangle & \langle \mathbf{v}_{\mu_2}, \mathbf{v}_{\mu_{j-1}}  \rangle & \dots &
\langle \mathbf{v}_{\mu_j}, \mathbf{v}_{\mu_{j-1}}  \rangle \\
\mathbf{v}_{\mu_1} & \mathbf{v}_{\mu_2} & \dots & \mathbf{v}_{\mu_j}  \end{vmatrix} \\
& = &  \frac{1}{\sqrt{D_{j-1} D_j}} \sum \limits^j_{i=1} (-1)^{j+i}  \mathbf{v}_{\mu_i} M_{j,i} \\
& = &  \frac{1}{\sqrt{D_{j-1} D_j}} \sum \limits^j_{i=1} (-1)^{j+i}  q(z \vert \mu_0, \Sigma_0)\dfrac{1}{2} \left[ (z-\mu_0)^T \Sigma_0^{-1} \right]_i M_{j,i} 
\end{eqnarray*}
where $D_0=1$ and $D_j$ is the Gram determinant for $j \geqslant 1$:
 $$D_j = \begin{vmatrix}
\langle \mathbf{v}_{\mu_1}, \mathbf{v}_{\mu_1} \rangle & \langle \mathbf{v}_{\mu_2}, \mathbf{v}_{\mu_1} \rangle & \dots & \langle \mathbf{v}_{\mu_j}, \mathbf{v}_{\mu_1} \rangle \\
\langle \mathbf{v}_{\mu_1}, \mathbf{v}_{\mu_2} \rangle & \langle \mathbf{v}_{\mu_2}, \mathbf{v}_{\mu_2} \rangle & \dots & \langle \mathbf{v}_{\mu_j}, \mathbf{v}_{\mu_2} \rangle \\
\vdots & \vdots & \ddots & \vdots \\
\langle \mathbf{v}_{\mu_1}, \mathbf{v}_{\mu_j} \rangle & \langle \mathbf{v}_{\mu_2}, \mathbf{v}_{\mu_j} \rangle & \dots &
\langle \mathbf{v}_{\mu_j}, \mathbf{v}_{\mu_j} \rangle \end{vmatrix}$$
$ M_{j,i} $ is a minor of: 
\begin{eqnarray*}
\begin{pmatrix}
\langle \mathbf{v}_{\mu_1}, \mathbf{v}_{\mu_1} \rangle & \langle \mathbf{v}_{\mu_2}, \mathbf{v}_{\mu_1} \rangle & \dots & \langle \mathbf{v}_{\mu_j}, \mathbf{v}_{\mu_1} \rangle \\
\langle \mathbf{v}_{\mu_1}, \mathbf{v}_{\mu_2} \rangle & \langle \mathbf{v}_{\mu_2}, \mathbf{v}_{\mu_2} \rangle & \dots & \langle \mathbf{v}_{\mu_j}, \mathbf{v}_{\mu_2} \rangle \\
\vdots & \vdots & \ddots & \vdots \\
\langle \mathbf{v}_{\mu_1}, \mathbf{v}_{\mu_{j-1}}  \rangle & \langle \mathbf{v}_{\mu_2}, \mathbf{v}_{\mu_{j-1}}  \rangle & \dots &
\langle \mathbf{v}_{\mu_j}, \mathbf{v}_{\mu_{j-1}}  \rangle \\
\mathbf{v}_{\mu_1} & \mathbf{v}_{\mu_2} & \dots & \mathbf{v}_{\mu_j}
 \end{pmatrix}
\end{eqnarray*}, 
obtained by taking the determinant of this matrix with row $j$ and column $i$ removed. To simplify the calculation, we could equivalently treat $ M_{j,i} $ as a minor of: 
\begin{eqnarray*}
A_j & = & \begin{pmatrix}
\langle \mathbf{v}_{\mu_1}, \mathbf{v}_{\mu_1} \rangle & \langle \mathbf{v}_{\mu_2}, \mathbf{v}_{\mu_1} \rangle & \dots & \langle \mathbf{v}_{\mu_j}, \mathbf{v}_{\mu_1} \rangle \\
\langle \mathbf{v}_{\mu_1}, \mathbf{v}_{\mu_2} \rangle & \langle \mathbf{v}_{\mu_2}, \mathbf{v}_{\mu_2} \rangle & \dots & \langle \mathbf{v}_{\mu_j}, \mathbf{v}_{\mu_2} \rangle \\
\vdots & \vdots & \ddots & \vdots \\
\langle \mathbf{v}_{\mu_1}, \mathbf{v}_{\mu_j}\rangle & \langle \mathbf{v}_{\mu_2}, \mathbf{v}_{\mu_j}\rangle & \dots &
\langle \mathbf{v}_{\mu_j}, \mathbf{v}_{\mu_j} \rangle \end{pmatrix} \\
\end{eqnarray*} 
since row $j$ is crossed out anyway. Notice that $A_j$ is the $j$th order leading principal submatrix of $ \dfrac{1}{4} \Sigma_0^{-1}$.

For basis
$\{ {v_l}_i \}_{i=1}^{\frac{D(D+1)}{2}}$, we have already obtained its Gram matrix $B$. We denote $B_j$ the $j$th order leading principal submatrix of $B$. The rest of the procedure is similar to deriving $\{ w_{\mu_i} \}_{i=1}^D$ from  $\{v_{\mu_i} \}_{i=1}^D$.

\paragraph{Updating the approximating distribution}
We have $$
 \overrightarrow{v_0}  =  \dfrac{\sum \limits_{i=1}^D \langle {w_\mu}_i, \sqrt{p_0}\rangle {w_\mu}_i + \sum \limits_{i=1}^{D(D+1)/2} \langle {w_l}_i, \sqrt{p_0}\rangle {w_l}_i}{\sqrt{1-\langle \theta_0, \sqrt{p_0}\rangle^2}}$$
Therefore, we use the following updates:
\begin{eqnarray*}
 \mu_i^{(t+1)} & = & \mu_i^{(t)}+\epsilon_{\alpha_i} \langle w_{\mu_i}, \sqrt{p_0}\rangle  / \sqrt{1-\langle \theta_0, \sqrt{p_0}\rangle^2} \quad  i = 1, \cdots , D\\
 l_i^{(t+1)} & = &l_i^{(t)}+\epsilon_{\beta_i} \langle {w_l}_i, \sqrt{p_0}\rangle  / \sqrt{1-\langle \theta_0, \sqrt{p_0}\rangle^2} \quad  i = 1, \cdots , \frac{D(D+1)}{2}
\end{eqnarray*} where $\epsilon_{\alpha_i}, \epsilon_{\beta_i} $ are stepsizes. As we can see, in order to update the parameters, all we need to calculate is $\langle w_{\mu_j}, \sqrt{p_0}\rangle$,  $\langle {w_l}_i, \sqrt{p_0} \rangle$ and $\langle \theta_0, \sqrt{p_0}\rangle$:
\begin{eqnarray*}
\langle w_{\mu_j}, \sqrt{p_0}\rangle & =  & \int \sqrt{p_0(z)}  \frac{1}{\sqrt{D_{j-1} D_j}} \sum \limits^j_{i=1} (-1)^{j+i}  q(z \vert \mu_0, \Sigma_0)\dfrac{1}{2} \left[(z-\mu_0)^T \Sigma_0^{-1} \right]_i M_{j,i} dz \\
 & =  & \frac{1}{\sqrt{D_{j-1} D_j}} \sum \limits^j_{i=1} (-1)^{j+i}  M_{j,i} E_{p(z \vert \mu_0, \Sigma_0)}\left( \frac{\sqrt{p_0(z)}}{q(z \vert \mu_0, \Sigma_0)}\frac{1}{2} \left[  (z-\mu_0)^T \Sigma_0^{-1} \right]_i \right) \\
\langle {w_l}_i, \sqrt{p_0} \rangle 
 & = & \dfrac{1}{\sqrt{E_{j-1}E_j}} \sum_{i=1}^j (-1)^{j+i} N_{j,i} E_{p(z \vert \mu_0, \Sigma_0)} \left( \dfrac{\sqrt{p_0(z)}}{q(z \vert \mu_0, \Sigma_0)} \left[ W_DU_DV_D \right]_i \right) \\
 \langle \theta_0, \sqrt{p_0}\rangle & = & E_{p(z \vert \mu_0, \Sigma_0)} \left( \dfrac{\sqrt{p_0(z)}}{q(z \vert \mu_0, \Sigma_0)} \right)
\end{eqnarray*}
Here, $E_j$ is the determinant of $B_j$ and $N_{j,i}$ is the minor of $B_j$. Expectations with respect to $p(z \vert \mu_0, \Sigma_0)$ can be approximated by the Monte Carlo method.

\section{Illustrative example: $t$-distribution}
\label{appd: tDist}
We now discuss the details for approximating $t(1)$ with a normal distribution. For a specific $\theta_0$, the basis of the tangent space $T_{\theta_0}\Theta$ is
\begin{eqnarray*}
v_{\mu} = \dfrac{\partial q}{\partial \mu} \Big \vert_{\mu = \mu_0, \sigma =\sigma_0} & = & q(x \vert \mu_0, \sigma^2_0)\dfrac{1}{2\sigma^2_0}(x-\mu_0) \\ 
v_{\sigma} = \dfrac{\partial q}{\partial \sigma} \Big \vert_{\mu = \mu_0, \sigma =\sigma_0} & = & q(x \vert \mu_0, \sigma_0^2)\left[ -\dfrac{1}{2}(\sigma_0)^{-1} + \dfrac{1}{2}\sigma_0^{-3}(x-\mu_0)^2 \right]\ \mbox{, where } -\infty < \sigma < \infty
\end{eqnarray*}
with the corresponding inner product
\begin{eqnarray*}
\langle v_{\mu},v_{\mu} \rangle & = &\int^{+\infty}_{-\infty} v_{\mu} v_{\mu}dx  = E_{q^2(x \vert \mu_0, \sigma_0^2)}(\dfrac{1}{2\sigma^2_0}(x-\mu_0) \dfrac{1}{2\sigma^2_0}(x-\mu_0) )=\dfrac{1}{4\sigma^2_0}\\ 
\langle v_{\sigma},v_{\sigma}\rangle & = &\int^{+\infty}_{-\infty} v_{\sigma}v_{\sigma}dx  =  E_{q^2(x \vert \mu_0, \sigma_0^2)}(\left[ -\dfrac{1}{2}(\sigma_0)^{-1} + \dfrac{1}{2}\sigma_0^{-3}(x-\mu_0)^2 \right]\left[ -\dfrac{1}{2}(\sigma_0)^{-1} + \dfrac{1}{2}\sigma_0^{-3}(x-\mu_0)^2 \right])= \dfrac{1}{2} \sigma_0^{-2}\\
\langle v_{\mu},v_{\sigma} \rangle & = &\int^{+\infty}_{-\infty} v_{\mu}v_{\sigma} dx  = E_{q^2(x \vert \mu_0, \sigma_0^2)}(\dfrac{1}{2\sigma^2_0}(x-\mu_0)\left[ -\dfrac{1}{2}(\sigma_0)^{-1} + \dfrac{1}{2}\sigma_0^{-3}(x-\mu_0)^2 \right]) = 0  
\end{eqnarray*}
Therefore, we can obtain an orthonormal basis as follows: 
\begin{eqnarray*}
w_{\mu} & = & \dfrac{v_{\mu}}{\Vert v_{\mu}\Vert} =  \dfrac{v_{\mu}}{\sqrt{\langle v_{\mu},v_{\mu} \rangle}} \\
& = & q(x \vert \mu_0, \sigma^2_0) \dfrac{1}{\sqrt{\sigma^2_0}}(x-\mu_0) \\
w_{\sigma} & = & \dfrac{v_{\sigma} }{\Vert v_{\sigma} \Vert} = \dfrac{v_{\sigma} }{\sqrt{\langle v_{\sigma},v_{\sigma}\rangle} }\\
& = &  q(x \vert \mu_0, \sigma^2_0) \dfrac{\sqrt{2}}{2} (\dfrac{(x-\mu_0)^2}{\sigma^2_0}-1)
\end{eqnarray*}
Given $\overrightarrow{v_0} = \dfrac{\langle w_{\mu}, \sqrt{p_0}\rangle w_{\mu}+\langle w_{\sigma}, \sqrt{p_0}\rangle w_{\sigma} }{\sqrt{1-\langle \theta_0, \sqrt{p_0}\rangle^2 }}$, we use the following updates:
\begin{eqnarray*}
\mu^{(t+1)} & = & \mu^{(t)}+\epsilon_{\alpha} \dfrac{\langle w_{\mu}, \sqrt{p_0}\rangle}{\sqrt{1-\langle \theta_0, \sqrt{p_0}\rangle^2 }} \\
{\sigma}^{(t+1)} & = & {\sigma}^{(t)}+\epsilon_{\beta} \dfrac{\langle w_{\sigma}, \sqrt{p_0}\rangle }{\sqrt{1-\langle \theta_0, \sqrt{p_0}\rangle^2 }}
\end{eqnarray*}
We calculate $\langle w_{\mu}, \sqrt{p_0}\rangle, \langle w_{\sigma}, \sqrt{p_0}\rangle,  \langle \theta_0, \sqrt{p_0}\rangle$ as follows
\begin{eqnarray*}
\langle w_{\mu}, \sqrt{p_0}\rangle  & = & \int^{+\infty}_{-\infty}w_{\mu} \sqrt{p_0} dx \\
& = & E_{q^2(x \vert \mu_0, \sigma_0^2)}(\dfrac{\pi^{-\frac{1}{2}}(1+x^2)^{-\frac{1}{2}} }{q(x \vert \mu_0, \sigma^2_0) } \dfrac{1}{\sqrt{\sigma^2_0}}(x-\mu_0) )\\
& = & \left(\dfrac{\pi}{2 \sigma_0^2} \right)^{-\frac{1}{4}} \dfrac{c_2}{\sqrt{\sigma_0^2}}\\
\langle w_{\sigma}, \sqrt{p_0}\rangle   & = & \int^{+\infty}_{-\infty}w_{\sigma} \sqrt{p_0} dx \\
& = & E_{q^2(x \vert \mu_0, \sigma_0^2)}(\dfrac{\pi^{-\frac{1}{2}}(1+x^2)^{-\frac{1}{2}} }{q(x \vert \mu_0, \sigma^2_0) } \dfrac{\sqrt{2}}{2} (\dfrac{(x-\mu_0)^2}{\sigma^2_0}-1))\\
& = &  \left(\dfrac{\pi}{2\sigma^2_0} \right)^{-\frac{1}{4}} (\dfrac{\sqrt{2}c_1}{2\sigma^2_0} - \dfrac{\sqrt{2}c_3}{2}) \\
\langle \theta_0, \sqrt{p_0}\rangle & = &  \int^{+\infty}_{-\infty}  q(x \vert \mu_0, \sigma_0^2)\sqrt{p_0} dx \\
& = & E_{q^2(x \vert \mu_0, \sigma_0^2)}(\dfrac{\pi^{-\frac{1}{2}}(1+x^2)^{-\frac{1}{2}} }{q(x \vert \mu_0, \sigma^2_0) })\\
 & = &   \left(\dfrac{\pi}{2\sigma^2_0} \right)^{-\frac{1}{4}} c_3 
\end{eqnarray*}
where
\begin{eqnarray*}
 c_1 & = &  E_{q_0^2} (\dfrac{(x-\mu_0)^2}{\sqrt{1+x^2}\exp(-\frac{1}{4\sigma^2_0}(x-\mu_0)^2)}) \\
 c_2 & = &  E_{q_0^2}  (\dfrac{x-\mu_0}{\sqrt{1+x^2}\exp(-\frac{1}{4\sigma^2_0}(x-\mu_0)^2)})  \\
 c_3 & = & E_{q_0^2}  (\dfrac{1}{\sqrt{1+x^2}\exp(-\frac{1}{4\sigma^2_0}(x-\mu_0)^2)} 
\end{eqnarray*}

%\paragraph{The difficulty in obtaining the exact solution}
%The geodesic $\gamma(t) \equiv (x^1(t),x^2(t))$ with initial position $\theta_0$ and initial velocity $\overrightarrow{v_0}$ can be obtained by solving:
%\[\begin{cases}
%\ddot{x}^1 = \dfrac{1}{\sigma^2}\dot{x}^1\dot{x}^2\\
%\ddot{x}^2 =  -\dot{x}^1\dot{x}^1+  \dfrac{1}{\sigma^2}\dot{x}^2\dot{x}^2
%\end{cases}\]
%, which doesn't have an explicit solution.
%\begin{eqnarray*}
%\mbox{Metric tensor: }  G & = & \begin{pmatrix}
%\dfrac{1}{\sigma^2} & 0 \\
%0 & \dfrac{1}{2 \sigma^4}  \\
%\end{pmatrix} \\
%\mbox{Inverse of G: } G^{-1} & = & \begin{pmatrix}
%\sigma^2 & 0 \\\bigsqcup
%0 & 2\sigma^4 \\
%\end{pmatrix} \\
%g_{11,2}  =  -\dfrac{1}{\sigma^4} & , & g_{22,2}  =  -\dfrac{1}{\sigma^6} \\
%\Gamma^1_{12} =\Gamma^1_{21}  =   -\dfrac{1 }{2\sigma^2} &, &  \Gamma^2_{11}  =1, \Gamma^2_{22}  =  -\dfrac{1}{\sigma^2}\\  
%\end{eqnarray*}
 
\clearpage
 
\bibliographystyle{plain} 
%\bibliography{refList}

\end{document}